\RequirePackage{ifpdf}
\documentclass[11pt,a4paper]{article}
\usepackage{amsmath,amssymb,epsfig,jheppubme}
\usepackage{multirow,longtable,enumerate,mathtools}
\usepackage{pdflscape, array}
\usepackage[raggedrightboxes]{ragged2e}
\usepackage{changepage}
\usepackage{bm}

\font\cmss=cmss12

\newcommand\half{\frac12}

\newcommand\bi{\begin{itemize}}
\newcommand\ei{\end{itemize}}

\newcommand\bea{\begin{eqnarray}}
\newcommand\eea{\end{eqnarray}}
\newcommand\be{\begin{equation}}
\newcommand\ee{\end{equation}}

\newcommand\btau{{\bar \tau}}

\newcommand\cN{{\cal N}}

\newcommand{\cO}{{\cal O}}

\newcommand\bchi{{\overline \chi}}

\newcommand\mA{{\mathsf A}}

\newcommand\mD{{\mathsf D}}
\newcommand\mE{{\mathsf E}}
\newcommand\mF{{\mathsf F}}
\newcommand\mG{{\mathsf G}}

\newcommand\sfrac[2]{{\textstyle\frac{#1}{#2}}}

\newcommand\ZZ{\hbox{Z\kern-.4emZ}}
\newcommand\sZZ{\hbox{\sevenfont Z\kern-.4emZ}}

\newcommand{\eref}[1]{Eq.\,(\ref{#1})}
\newcommand{\Comment}[1]{{}}

\newcommand\Tstrut{\rule{0pt}{3.6ex}}

\def\IB{\relax{\rm I\kern-.18em B}}
\def\IC{{\relax\hbox{\kern.3em{\cmss I}$\kern-.4em{\rm C}$}}}
\def\ID{\relax{\rm I\kern-.18em D}}
\def\IE{\relax{\rm I\kern-.18em E}}
\def\IF{\relax{\rm I\kern-.18em F}}
\def\II{\relax{\rm I\kern-.18em I}}
\def\IZ{\mathbb{Z}}
\def\Id{\relax{1\kern-.32em 1}}
\def\IG{\relax\hbox{$\inbar\kern-.3em{\rm G}$}}
\def\IR{\relax{\rm I\kern-.18em R}}

\newcommand{\tMmax}{{\tilde M}_{\rm max}}
\newcommand{\tMmin}{{\tilde M}_{\rm min}}

\title{Signs, growth and admissibility of quasi-characters and the holomorphic modular bootstrap for RCFT} 
\author[\,a,b]{Arpit Das }
\author[\,c]{and Sunil Mukhi \footnote{Adjunct Professor, ICTS Bengaluru.}} 

\affiliation[a]{School of Mathematics, University of Edinburgh,\\ Edinburgh EH9 3FD, U.K.}
\affiliation[b]{Higgs Centre for Theoretical Physics,
University of Edinburgh,\\ Edinburgh EH8 9YL, U.K.}
\affiliation[c]{Indian Institute of Science Education and Research,\\ Homi Bhabha Rd, Pashan, Pune 411 008, India}

\emailAdd{arpit.das@ed.ac.uk}
\emailAdd{sunil.mukhi@gmail.com}

\abstract{Rational conformal field theories in 2d have partition functions built from holomorphic characters, whose classification can be addressed via the holomorphic modular bootstrap. This is facilitated by a special basis of ``quasi-characters'' that has been completely classified for rank-2. Suitably combining these to form admissible characters with non-negative integral coefficients $a_n$ depends crucially on the signs and growth of the quasi-character coefficients.  We use Frobenius recursion relations for Modular Linear Differential Equations to estimate the growth with $c$ of these coefficients in the region $n\sim\frac{c}{12}$ that is inaccessible to Cardy asymptotics, and to prove rigorously that they have alternating signs that stabilise to a fixed sign at this order. This provides a practical path to obtain candidate RCFT partition functions at arbitrary Wronskian index.}

\preprint{}

\begin{document}

\maketitle

\parindent=0pt
\advance\parskip by 3pt

\section{Introduction and Motivation}

The classification of Rational Conformal Field Theories in 2d (RCFT) has seen significant progress in the last decade \cite{Gaberdiel:2016zke, Mason:2018, connor2018classification, Mukhi:2022bte, Rayhaun:2023pgc, Das:2022slz, Das:2022uoe}. A key ingredient is the classification of possible characters that are vector-valued modular functions under $\rm SL(2,\mathbb{Z})$ and also physically admissible, having integral coefficients $a_n$ in a $q$-series \footnote{We are using the notation vector-valued modular function for a vector-valued modular form under $\rm SL(2,\mathbb{Z})$ with weight zero. By a slight abuse of notation at times we will use the word fucntion or form to mean the same object.}. These two criteria ensure a modular-invariant partition function as well as the possibility of interpreting it as a generating function for the degeneracies of states of a (Euclidean) quantum field theory. The space of holomorphic $q$-series satisfying these criteria is therefore a superset of the (still largely un-understood) space of RCFT. 

The partition function $Z(\tau,\btau)$ of an RCFT can be expanded over a finite set of (generalised) characters $\chi_i(\tau)$ that are holomorphic in the interior of the torus moduli space parametrised by $\tau$:
\be
Z(\tau,\btau)=\sum_{i=0}^{p-1}\chi_i(\tau)M_{ij}\bchi_{j}(\btau)
\ee
The characters count descendants under the full chiral algebra, that is left unspecified. Let $r$, called the rank, denote the number of primary fields $\phi_i$ of the theory. There is one character for each primary field, though they are not necessarily linearly independent -- for example, a field $\phi_i$ and its complex conjugate $\phi_{\bar i}$ will have the same character, even though they can be distinct as primaries. Hence, the total number $p$ of characters satisfies $p\le r$.

The full set of characters (counting repeated ones) makes up a vector-valued modular form (VVMF), which transforms under $\rm SL(2,\mathbb{Z})$ as:
\be
\chi_i\left(\frac{a\tau+b}{c\tau+d}\right)
=\sum_j\varrho_{ij}\begin{pmatrix} a& b\\ c&d \end{pmatrix}\chi_j,\quad \begin{pmatrix} a& b\\ c&d \end{pmatrix}\in \hbox{\rm SL(2,$\mathbb{Z}$)}
\label{VVMFtrans}
\ee
where the modular representation matrix $\varrho$ provides an $r$-dimensional unitary representation of $\rm SL(2,\mathbb{Z})$ and satisfies $\varrho^\dagger M \varrho = M$ \footnote{As explained in \cite{Mathur:1988na, Mathur:1988gt}, \eref{VVMFtrans} remains true for the -- in general smaller -- set of $p$ independent characters, but for example the matrix $S=\varrho\begin{pmatrix} 0 & -1\\ 1 & ~0 \end{pmatrix}$ only satisfies $S^2=1$ and is no longer unitary. Instead, it preserves a diagonal matrix ${\rm diag}(1,m)$ where $m$ is the multiplicity of the repeated primaries. A procedure to relate the unitary $r\times r$ and the non-unitary $p\times p$ matrices was described in \cite{Mathur:1988gt}.}.

The components $\chi_i$ of the VVMF have a series expansion in terms of $q=e^{2\pi i\tau}$ of the form:
\be
\chi_i(\tau)=q^{\alpha_i}\sum a_{i,n}q^n
\ee
A VVMF will be called {\em admissible} if $a_{0,0}=1$ and $a_{i,n}\in \IZ^+$ for all $i,n$. This is a necessary but not sufficient condition for them to correspond to a sum over descendants under a chiral algebra. 

Historically, the classification of admissible VVMFs was addressed using two distinct approaches: Modular Linear Differential Equations \cite{Mathur:1988na, Mathur:1988gt} and Bantay-Gannon theory \cite{Bantay:2005vk, Bantay:2007zz}. The former approach works particularly well in low rank and low Wronskian index $\ell$, and registered a striking success by completely classifying all rank-2 RCFT with $\ell=0$ \footnote{Three decades later, this classification was rigorously shown in \cite{Mason:2018} to be complete. The classification for rank 3 and $\ell=0$ was completed in \cite{Das:2022uoe}.}. In this approach, no a priori data regarding modular representations or exponents is required. On the other hand, Bantay-Gannon theory starts with representation theory, namely the classification of modular data -- a representation of $\rm SL(2,\mathbb{Z})$ specified by its $S$ and $T$ matrices -- and with some additional physical inputs, constructs VVMFs that transform in that representation. Hybrid approaches have also been considered, for example in \cite{Bantay:2010uy, Cheng:2020srs, Govindarajan:2025rgh, Govindarajan:2025jlq}.

In \cite{Chandra:2018pjq} it was shown that for the two-character case, it is actually enough to find all solutions of the simplest MLDEs, namely the ones with $\ell=0,2,4$, subject to certain integrality properties and take their linear combinations to generate $\ell=6r, 6r+2, 6r+4$ for all $r>0$. These solutions are classified, and are referred to as quasi-characters. The price one pays in this approach is that most quasi-characters are not themselves admissible because some of their coefficients are typically negative integers. However, well-chosen  linear combinations of quasi-characters can be admissible. Moreover some of them were shown in \cite{Chandra:2018ezv} to describe genuine RCFTs. Thus the problem of classifying admissible rank-2 VVMFs becomes one of determining which linear combinations of quasi-characters are admissible. 
This, in turn, requires us to understand the nature of the negative coefficients of quasi-characters, the main goal of the present work. 

Rank-2 quasi-characters  were adapted in \cite{Chandra:2018pjq} from an infinite family of solutions to a Modular Linear Differential Equation (MLDE) studied in \cite{KZ, KK, Kaneko:On}, motivated by certain mathematical  applications. As MLDE solutions, they are characterised by a weaker requirement than admissibility: integrality of the coefficients $a_{i,n}$ is still required, but they need not all be positive. Also we no longer insist on $a_{0,0}=1$, instead there only has to exist {\em some} overall normalisation that makes the coefficients integral. Thus (barring a small number of cases), quasi-characters are non-admissible on their own and do not describe CFT characters. However, as was argued in \cite{Chandra:2018pjq}, they form a linear basis for admissible characters of arbitrary Wronskian index. 

\Comment{

Although we will focus on VVMFs that have precisely two independent components, $p=2$ in the above notation, it should be noted that these can arise in the context of RCFT with $r= 2,3$ or $4$ independent primaries. In the latter two cases there are symmetries due to which two or even three primaries have the same character \footnote{While the actual rank is relevant for the classification of CFTs, particularly when associating them to a Modular Tensor Category (MTC) \cite{Rowell2009}, only the number $p$ of independent characters is relevant when solving MLDEs.}. A complete set of quasi-characters for $p=2$ was constructed in  \cite{Chandra:2018pjq}. 

}

In \cite{Chandra:2018pjq} several properties of quasi-characters were conjectured based on explicit examples. The properties include a pattern of alternating signs of coefficients in the $q$-series that eventually stabilise to a definite sign at a term $q^n$ where $n\sim \frac{c}{12}$. In some cases there is no alternation and in others, the alternation follows two different patterns -- one up to $n\sim \frac{c}{24}$ and the other from $n\sim \frac{c}{24}$ to $n\sim\frac{c}{12}$. However, no derivations of these properties were provided in \cite{Chandra:2018pjq}. 

Our goal here is to provide rigorous proofs of these properties for the $\ell=0$ family of quasi-characters, which will be defined below. On the way, we summarise the results of \cite{Chandra:2018pjq} in a different -- and hopefully more useful -- form and expand on them in some detail, filling in several gaps in the original results and correcting a few of the conclusions. We also find estimates of the growth of $q$-coefficients in the interesting regions where the sign stabilises. These are important in order to understand exactly how admissible characters are built out of quasi-characters. 

\Comment{

Another approach to the classification of VVMF was proposed by Bantay and Gannon \cite{Bantay:2005vk, Bantay:2007zz, Gannon:2013jua}. This approach is powerful and gives a uniform construction for any rank as long as the modular representation is known. However in the quasi-character approach we work with solutions of an MLDE, unlike the Bantay-Gannon basis. It is also worth noting that in this context the modular representation itself is one of the outputs from the MLDE \cite{Mathur:1988gt} (see also \cite{Das:2023qns}). The limitation so far is that quasi-characters have only been fully classified in the two-character case (a partial attempt for three characters was made in \cite{Mukhi:2020gnj}).
}

We mention here that the classification of admissible VVMFs is a tractable problem for which it is conceivable that a complete set of rules exists at low rank. By contrast, even at low rank the classification of all RCFT is impossibly hard. This fact has long been known in rank-1 (meromorphic) CFTs \cite{Schellekens:1992db} and in recent years has been explicitly demonstrated for rank $>1$ as well \cite{Harvey:2018rdc, Chandra:2018pjq, Chandra:2018ezv, Das:2022uoe}. The existence of a CFT for a given set of admissible characters is a separate (but important) question, on which significant progress has been made \cite{Gaberdiel:2016zke, franc2020classification, Duan:2022ltz,  Mukhi:2022bte, Das:2022uoe, Rayhaun:2023pgc, Moller:2024plb, Moller:2024xtt}, but it will not be addressed in the present work.

\section{Survey of quasi-characters}

\subsection{Review of quasi-characters}

\label{qcreview}

Quasi-characters are non-admissible but integral solutions to Modular Linear Differential Equations (MLDE). The most general second-order MLDE has the form:
\be
\Big(D_\tau^2+\sfrac{\ell}{6}\phi_2(\tau) D_\tau+ \phi_4(\tau)\Big)\chi=0
\label{gentwo}
\ee
where:
\be
D_\tau\equiv\frac{1}{2\pi i}\frac{d}{d\tau}-\frac{w}{12}E_2(\tau)
\ee
when acting on a weight-$w$ modular function and $E_2(\tau)$ is the second Eisenstein series, which transforms as a connection on moduli space. The function $\phi_2,\phi_4$ are modular of weight 2 and 4 respectively under $\rm SL(2,\mathbb{Z})$. In general they are meromorphic and can have $\sfrac{\ell}{6}$ poles in $\tau$, where the fractional number comes from the singular points in torus moduli space (see \cite{Chandra:2018pjq} for details) and $\ell$ is an integer called the Wronskian index. The solutions $\chi$ of the equation are generically holomorphic in $\tau$, except at $\tau\to i\infty$ where they can diverge.

The solutions of \eref{gentwo} have an expansion of the form:
\be
\chi_i(q)=q^{\alpha_i}\sum_{n=0}^\infty a_{i,n}q^n
\ee
where $i=0,1$ for the identity and non-identity character respectively. The $\alpha_i$ are called the exponents of the solution.
They satisfy the valence formula:
\be
\alpha_0+\alpha_1=\frac{1-\ell}{6}
\label{valence}
\ee
It is useful to label the exponents by $\alpha_0=-\frac{c}{24}, \alpha_1=-\frac{c}{24}+h$. In the case where MLDE solutions describe a CFT, $c$ and $h$ are the (chiral) central charge and conformal dimension for that CFT, otherwise this is just a choice of notation. In terms of $c,h$, the valence formula becomes:
\be
\ell=1-6h+\frac{c}{2}
\label{valence.ch}
\ee
For the special case of second-order MLDE, it has been shown that the Wronskian index is even \cite{Naculich:1988xv} (see also \cite{Das:2023qns}). 

The central result of \cite{Chandra:2018pjq} is that quasi-characters with $\ell=0,2,4$ provide a basis for all admissible VVMFs with Wronskian index $\ell=6n, 6n+2, 6n+4$ respectively, where $n$ is any non-negative integer. Given that MLDEs with $\ell<6$ are relatively easy to solve explicitly, this is a useful simplification and in principle permits a complete classification of admissible characters.

For $\ell=0$, the quasi-characters are solutions to the MMS equation \cite{Mathur:1988na}:
\be
(D_\tau^2+\mu E_4)\chi=0
\label{MMSeq}
\ee
Here $\mu$ is a real parameter, related to the central charge of the solution by:
\be
\mu=-\frac{c(c+4)}{576}
\label{muc}
\ee
The construction of these quasi-characters was inspired by explicit results in the theory of second-order Modular Linear Differential Equations \cite{KZ, KK, Kaneko:On}. It should be noted that these references only address the case of vanishing Wronskian index $\ell=0$, which in the mathematical literature is sometimes called the ``non-zero Wronskian condition'' (because $\ell=0$ means that the Wronskian determinant, which appears in the denominator of $\phi_2$ and $\phi_4$, is everywhere non-vanishing).

For Wronskian index $\ell=2$, the quasi-character solutions were again constructed in \cite{Chandra:2018pjq}. Next, it was argued that all $\ell=4$ quasi-characters take the form $\chi_{\mE_{8,1}}$ multiplied by the $\ell=0$ quasi-characters (see also \cite{Das:2023qns}). This means that using only solutions of the MLDEs with $\ell=0$ and 2, we can construct all admissible characters for any allowed $\ell$. This is a very powerful result.

In the present work we focus exclusively on $\ell=0$ quasi-characters, leaving the more complicated $\ell=2$ family for future work \cite{upDM}. Let us now provide the details of the quasi-character solutions. The MMS equation has 10 apparently admissible solutions that were classified in \cite{Mathur:1988na}. Of these, two cannot correspond to unitary CFTs while the remaining 8 were identified with the WZW models for $\mA_{1,1},\mA_{2,1}, \mG_{2,1},\mD_{4,1}, \mF_{4,1},\mE_{6,1},\mE_{7,1},\mE_{8,1}$. The last one is an outlier in that its identity character gives a rank-1 (meromorphic) CFT while its second character is spurious. The remaining 7 are pairs of potentially unitary admissible characters \footnote{Three decades after they were classified, it was proved in the mathematical literature that this identification is complete and each of these admissible characters corresponds to a unique CFT \cite{Mason:2018}.}. 

Now as shown in \cite{Chandra:2018pjq}, \eref{MMSeq} admits infinitely many additional solutions with the following properties: (i) for an appropriate choice of normalisation, the coefficients $a_n$ of the $q$-series are integers, (ii) one of the two solutions has coefficients that alternate in sign up to $n\sim \frac{c}{12}$ (more precisely $n=2M$, where $M$ is defined in \eref{cqcell.0} below) and then stabilise to positive signs for $M>0$ and to negative signs for $M<0$, (iii) the other solution has all positive coefficients, (iv) suitable linear combinations of quasi-characters are admissible characters with $\ell=6n$ and all such admissible characters can be constructed in this way. Subsequently in \cite{Chandra:2018ezv} the characters of a few actual CFTs were exhibited that are described by these solutions.

Whenever a solution with oscillatory signs stabilises to a positive sign, we refer to it as ``Type I'' and when it stabilises to a negative sign, as ``Type II''.

The infinite set of quasi-characters generalises the 7 admissible solutions of \eref{MMSeq} in a precise way \footnote{Our presentation of the list here is somewhat different from that in \cite{Chandra:2018pjq}. Readers interested in knowing how the two presentations are related may consult Appendix \ref{AppA}.}. We label their central charges by:
\be
c=24M+j
\label{cqcell.0}
\ee
where:
\be
j=1,2,\frac{14}{5},4, \frac{26}{5},6,7
\label{jvalell.0}
\ee
From the valence formula \eref{valence.ch} and $\ell=0$ we find that $h=2M+\frac{j+2}{12}$ for these quasi-characters. Thus the exponents are \footnote{Note that under $c\leftrightarrow -c-4$ or, equivalently, $24M+j\leftrightarrow -(24M+j+4)$ in \eref{exps.0}, we have the exchange of exponents, that is: $\alpha_0\leftrightarrow\alpha_1$. This also leads to exchange of the associated characters.}:
\be
\begin{split}
\alpha_0 &= -M-\frac{j}{24}\\[2mm]
\alpha_1 &= M+\frac{j+4}{24}
\end{split}
\label{exps.0}
\ee

\Comment{
\begin{table}
\centering
\begin{tabular}{|c|c|}
\hline
\Tstrut $\ell=0$ Series & Central charges\\[2mm]
\hline
\Tstrut  $\mG_2$ & $24M+\frac{14}{5}$\\[2mm]
 $\mF_4$ & $24M+\frac{26}{5}$\\[2mm]
\hline
\Tstrut  $\mA_1$ & $24M+1$\\[2mm]
 $\mE_7$ & $24M+7$\\[2mm]
\hline
\Tstrut  $\mA_2$ & $24M+2$\\[2mm]
 $\mE_6$ & $24M+6$\\[2mm]
\hline
\Tstrut $\mD_4$ & $24M+4$\\[2mm]
\hline
\end{tabular}
\caption{Quasi-character sub-series: $\ell=0$}
\label{qcell.0}
\end{table}
}

Since the MLDE with $\ell=0$ depends on just one real parameter, specifying $c$ is sufficient to fix that parameter and hence the equation and its solution. The properties listed above can thus be explicitly verified for any fixed small value of $M$. It is also important to emphasise that since every quasi-character has a different $c$, it solves the same MLDE but with a different value of the parameter $\mu$ in each case, determined by \eref{muc}. 

To illustrate the behaviour described above, in Appendix \ref{explicitlow} we tabulate the first few coefficients for the $\mA_{1,1}$ series $-3\le M\le 3$. For $M=0$ we have $c=j$, which is precisely the set of central charges of the ``MMS theories'' \cite{Mathur:1988na}. These are the only admissible solutions. We may then apply their labels  $\mA_{1}, \mA_{2}, \mG_{2},\mD_{4},\mF_{4},\mE_{6},\mE_{7}$ \footnote{Since the Kac-Moody algebras all arise at level 1, it is sufficient to label them by the corresponding  finite-dimensional Lie sub-algebra.} to the corresponding families of quasi-characters \footnote{These labels also describe the Modular Tensor Category to which the MMS solutions can be associated.}.

\subsection{Frobenius recursion relation}

\label{Frobrecursubsec}

The Modular Linear Differential Equation \eref{MMSeq} which determines the quasi-characters can be solved by the Frobenius method in terms of the Eisenstein series coefficients and the central charge. The solution given here will be used later on for approximations as well as proofs by induction. The relevant Eisenstein series appearing in this equation are given by:
\be
\begin{split}
E_2(\tau)&=1-24\sum_{i=1}^\infty \sigma_1(i)q^i\\
E_4(\tau) &= 1+240 \sum_{i=1}^\infty \sigma_3(i)q^i
\end{split}
\ee
where $\sigma_m(i)\equiv\sum_{d|i}d^m$ is the generalised divisor function. The resulting recursion relation for the identity character is \cite{Mathur:1988na}:
\be
a_{0,n}=\frac{\sum_{i=1}^n a_{0,n-i}\left(-4 \big(n-i-\frac{c}{24}\big)\sigma_1(i)+\frac{5c(c+4)}{12}\sigma_3(i)\right)}{n(n-\frac{c+2}{12})}
\label{frobrecur.l0c0}
\ee

For the non-identity character we have instead:
\be
a_{1,n}=\frac{\sum_{i=1}^n a_{1,n-i}\left(-4 \big(n-i+\frac{c+4}{24}\big)\sigma_1(i)+\frac{5c(c+4)}{12}\sigma_3(i)\right)}{n(n+\frac{c+2}{12})}
\label{frobrecur.l0c1}
\ee

Now we use the parametrisation of $c$ in Eqs (\ref{cqcell.0}), (\ref{jvalell.0}) and re-write the Frobenius recursions as:
\be
a_n=\sum_{i=1}^n f_M(n,i)\,a_{n-i}
\label{recurf}
\ee
where the $f_M(n,i)$ are:
\be
\begin{split}
f_M(n,i) &\equiv \frac{-4 \big(n-i-M-\frac{j}{24}\big)\sigma_1(i)+\frac{5(24M+j)(24M+j+4)}{12}\sigma_3(i)}{n(n-2M-\frac{j+2}{12})} \qquad\qquad \hbox{(identity)}\\
f_M(n,i) &\equiv \frac{-4 \big(n-i+M+\frac{j+4}{24}\big)\sigma_1(i)+\frac{5(24M+j)(24M+j+4)}{12}\sigma_3(i)}{n(n+2M+\frac{j+2}{12})}\qquad \hbox{(non-identity)}
\end{split}
\label{fMni.0}
\ee

\subsection{Modular transformations}

It was noted above that with the quasi-characters being labelled by $c=24M+j$, the $M=0$ member of each series is in the MMS family \cite{Mathur:1988na}. This member has $c=j$ and its modular $S$ matrix was identified in \cite{Mathur:1988gt}:
\be
\begin{split}
j=\frac{14}{5},\frac{26}{5}:~~ S&=\sqrt{2}\begin{pmatrix}
\frac{1}{\sqrt{5+\sqrt{5}}} & ~\frac{1}{\sqrt{5 -\sqrt{5}}}\\
\frac{1}{\sqrt{5 -\sqrt{5}}} & -\frac{1}{\sqrt{5 + \sqrt{5}}}
\end{pmatrix}\\[2mm]
j=1,7 :~~ S&=\frac{1}{\sqrt2}\begin{pmatrix}
1 & ~1\\
1 & -1
\end{pmatrix}\\[2mm]
j=2,6:~~ S&=\frac{1}{\sqrt3}\begin{pmatrix}
1 & ~2\\
1 & -1
\end{pmatrix}\\
j=4 :~~ S&=~~\,\frac{1}{2}\begin{pmatrix}
1 & ~3\\[2mm]
1 & -1
\end{pmatrix}
\end{split}
\label{AllS}
\ee
In the first two lines, the given $S$-matrix is real, symmetric and unitary (hence it squares to 1). In the remaining two lines it is real but neither symmetric nor unitary, but still squares to 1. In the latter two cases it is not the true $S$-matrix. As explained in \cite{Mathur:1988gt}, in these cases one has more than two primary fields (hence they correspond to rank $>2$ MTC). In fact the $\mA_2, \mE_6$ series have three primaries, with two of them corresponding to complex conjugates of each other and thus having the same character, while the $\mD_4$ series has four primaries of which three (the $8_{\rm v},8_{\rm c},8_{\rm s}$ representations of $\mD_4$) are related by triality and thus have the same character. Thus the true $S$-matrices should have rank 3 and 4 respectively. The ones listed above then follow by ``collapsing'' the rows and columns corresponding to the same characters. The true $S$-matrices for these cases were constructed in \cite{Mathur:1988gt}. They coincide with the known $S$-matrices for the corresponding MTC \cite{Rowell2009}, however we do not need them as the collapsed matrices are sufficient for our purpose here.

Now although the above matrices are obtained for the admissible characters for $M=0$, they automatically describe the transformation of all quasi-characters in the corresponding family, because modular properties repeat whenever $c$ jumps by 24. Thus \eref{AllS} gives the (collapsed) $S$-matrix for each quasi-character series.

\subsection{Combining quasi-characters}

The utility of quasi-characters is that they form a convenient basis for constructing admissible characters. The basic idea is that, since all quasi-characters in a sub-series transform in the same way under $\rm SL(2,\mathbb{Z})$, we can add them to construct more general VVMFs. Each quasi-character considered here solves  \eref{MMSeq} but, as emphasised above, it does so for a different value of the parameter $\mu$. Hence the sum does {\em not} satisfy the same equation. Rather, with each successive term in the sum one finds that $\ell$ jumps by 6 units. We will show this below. In this way we can generate admissible characters with all $\ell=6r$ for any positive integer $r$. 

A generic linear combination of quasi-characters in any one of the sub-series with $c=24M+j$ (for fixed $j$) takes the form:
\be
\chi_i(\tau) =\sum_{M=M_{\rm min}}^{M_{\rm max}} \cN_M\, \chi_i^{[24M+j]}(\tau)
\label{qcsum.0}
\ee
where $j$ is given by \eref{jvalell.0}. Let us determine the central charge and conformal dimension of the sum. We use \eref{exps.0} for the exponents of the quasi-characters. Then the structure of the above sum is:
\be
\begin{split}
\chi_0(\tau) &= \cN_{M_{\rm max}}q^{-M_{\rm max}-\frac{j}{24}}+\cN_{M_{\rm max}-1}q^{-M_{\rm max}+1-\frac{j}{24}}+\cdots
+\cN_{M_{\rm min}}q^{-M_{\rm min}-\frac{j}{24}}\\
\chi_1(\tau) &= \cN_{M_{\rm max}}q^{M_{\rm max}+\frac{j+4}{24}}+\cN_{M_{\rm max}-1}q^{M_{\rm max}-1+\frac{j+4}{24}}+\cdots+
\cN_{M_{\rm min}}q^{M_{\rm min}+\frac{j+4}{24}}
\end{split}
\label{gensum.0}
\ee
As long as $M_{\rm max} > M_{\rm min}$ the ``most singular'' \footnote{This just means the term with the lowest power of $q$, though the term need not be singular in the sense of diverging as $q\to 0$.} term in the first line is the first term, while in the second line the most singular term is the last term. Thus we have:
\be
\alpha_0 = -M_{\rm max}-\frac{j}{24}, ~\alpha_1=M_{\rm min}+\frac{j+4}{24}
\ee
It follows that the sum has $\alpha_0+\alpha_1= -M_{\rm max}+ M_{\rm min}+\frac16$. From \eref{valence} we then find that:
\be
\ell=6(M_{\rm max}-M_{\rm min})
\label{ellsum.0}
\ee
In this way by adding increasing numbers of quasi-characters we can get $\ell$ to be an arbitrary multiple of 6.

From \eref{gensum.0} we can now read off the $(c,h)$ values of the sum of quasi-characters, to find:
\be
\begin{split}
c &= 24M_{\rm max}+j\\
h &= M_{\rm max}+M_{\rm min}+\frac{j+2}{12}
\end{split}
\label{chofsum.0}
\ee
Notice that if $M_{\rm max}+M_{\rm min}+\frac{j+2}{12}<0$ then $h$ is negative and we necessarily have a non-unitary theory. Now  \eref{jvalell.0} implies that $\frac14\le \frac{j+2}{12}\le \frac34$, so for any positive $M_{\rm max}$ the largest allowed number of terms in the sum consistent with unitarity is attained by choosing $M_{\rm min}=-M_{\rm max}$. Also the only case for which $h<1$ is $M_{\rm max}=M_{\rm min}=0$. With two characters and $h$ non-integer, the only primary in the full non-chiral CFT (if one exists) has dimensions $(h,h)$. Hence all cases with $M_{\rm max}\ne 0$, or more precisely the CFTs among them, correspond to what have been called ``perfect metals'' in \cite{Plamadeala:2014roa} --  such CFTs have no relevant deformations.

Another important point is that if $M_{\rm min}\ge 1$, one cannot have admissibility. This follows from the fact that in the entire sum \eref{qcsum.0} only the $M=0$ term is admissible by itself and without it all the negative terms cannot be cancelled.

Thus we have the following results:

\bi
\item
The central charge of the sum is entirely determined by $M_{\rm max}$.

\item
The conformal dimension $h$ of the sum, for fixed $M_{\rm max}$, can vary from $\frac{j+2}{12}$ to $2M_{\rm max}+\frac{j+2}{12}$ depending on the choice of $M_{\rm min}$. 

\item
The allowed Wronskian index $\ell$ for a given $M_{\rm max}$ satisfies $6M_{\rm max}\le \ell \le 12 M_{\rm max}$. 

\item
When $M_{\rm max}<0$ we cannot have a unitary theory (in fact we cannot even have admissible characters).

\ei

Since $M_{\rm max}$ determines the central charge, we get an important result for all unitary 2d RCFTs with two characters:  \\[-2mm]

{\em When $\ell=0$ mod 6, the allowed central charges for such CFT are $c=24M+j$ with $M$ a non-negative integer and $j\in \{1,2,\frac{14}{5},4,\frac{26}{5},6,7\}$. The maximum Wronskian index for such a theory is $\ell=\frac{c-j}{2}$.}\\[-2mm]

The first of the above results of course follows also from the well-known (many-to-one) identification of RCFT with MTC, but the second one is also useful.

\subsection{Admissible characters from quasi-characters}

Now we look more closely at admissibility and provide two examples that illustrate the key issues. The points that these examples will highlight are: (i) quasi-characters can be used to construct admissible characters, (ii) the signs of the coefficients of the quasi-characters play an essential role in this construction.

Our first example was already presented in sub-section 5.2 of \cite{Chandra:2018pjq} but we display it here because it exhibits an important feature that may not be well-recognised (also in \cite{Chandra:2018pjq} it was presented in a different notation from the more convenient one used here). We consider the $\mG_2$ family, whose central charges are $c=24M+\frac{14}{5}$. Choosing $\tMmax=1, \tMmin=0$, solving the MLDE \eref{MMSeq} and taking the linear combination in \eref{qcsum.0}, we get:
\be
\begin{split}
\chi_0 &= q^{-\frac{67}{60}}\Big(7\cN_1+(-1742\cN_1+\cN_0)\,q+
(722729\cN_1+14 \cN_0)\,q^2+\ldots
\Big)\\
\chi_1 &= q^{\frac{17}{60}}\Big(7\cN_0 + (34\cN_0 + 11884326\cN_1)q + (119\cN_0+1184238132\cN_1)q^2 + \ldots\Big)
\end{split}
\ee

We see that the identity character has just one potentially negative term, as expected from Table \ref{A1signs.0}. Thus as long as $\cN_0,\cN_1\ge 0$, all the remaining terms from order $q^2$ onwards will be non-negative. Under the same conditions, the non-identity character has all coefficients non-negative. Now to make the $\cO(q)$ term non-negative we must impose $\cN_0\ge 1742 \cN_1$. However this is not sufficient. The leading term of the identity character is $7\cN_1$, while for admissibility it needs to be 1. Choosing $\cN_1=1$ and $\cN_0=1742+7n$ for any non-negative integer $n$ gives: 
\be
\chi_0 = q^{-\frac{67}{60}}\Big(7+7nq+
(747117+98n)q^2+\cdots\Big)
\ee
Now we find that all coefficients to this order are multiples of 7, and remarkably this fact persists to all the orders we have calculated. This was not guaranteed! We can now re-scale $\chi_0$ by dropping an overall 7, to get:
\be
\chi_0 = (1 + nq+ (106731+ 14n) q^2 + (19112822+ 294n)q^3 + (1053497615+1960n) q^4+\cdots
\ee
With these choices of $\cN_1,\cN_0$, the non-identity is easily verified to be admissible.  Thus we have generated an infinite family of admissible characters generated by adding two quasi-characters. This family has $(c,h)=(\frac{134}{5}, \frac75)$ and $\ell=6$.

For the second example, briefly discussed in \cite{Das:2023qns}, we combine three quasi-characters in the $\mA_1$ family by choosing $j=1, M_{\rm max}=1,M_{\rm min}=-1$ in \eref{qcsum.0}. Then we get:
\be
\begin{split}
\chi_0 & =q^{-\frac{25}{24}}\Big(\cN_1 + (-245\cN_1 + \cN_0) q + (142640\cN_1 + 3\,\cN_0 + 26752\, \cN_{-1}) q^2 \\
&\qquad + (18615395\cN_1 + 4\, \cN_0 + 1734016\, \cN_{-1}) q^3\\
&\qquad + (837384535\cN_1 + 7\,\cN_0 + 46091264\, \cN_{-1}) q^4 + \ldots\Big)\\
\chi_1 &= q^{-\frac{19}{24}} \Big(\cN_{-1} + (2\, \cN_0 - 247\, \cN_{-1}) q \\
&\qquad + (565760\cN_1 + 2\, \cN_0 - 86241\, \cN_{-1}) q^2 \\
&\qquad + (51745280\cN_1 + 6\, \cN_0 - 4182736\, \cN_{-1}) q^3\\
&\qquad + (1965207040\cN_1 + 8\, \cN_0 - 96220123\, \cN_{-1}) q^4 + \ldots\Big) 
\end{split}
\label{ell12char}
\ee

\begin{figure}[h!]
\centering{
\includegraphics[height=8cm]{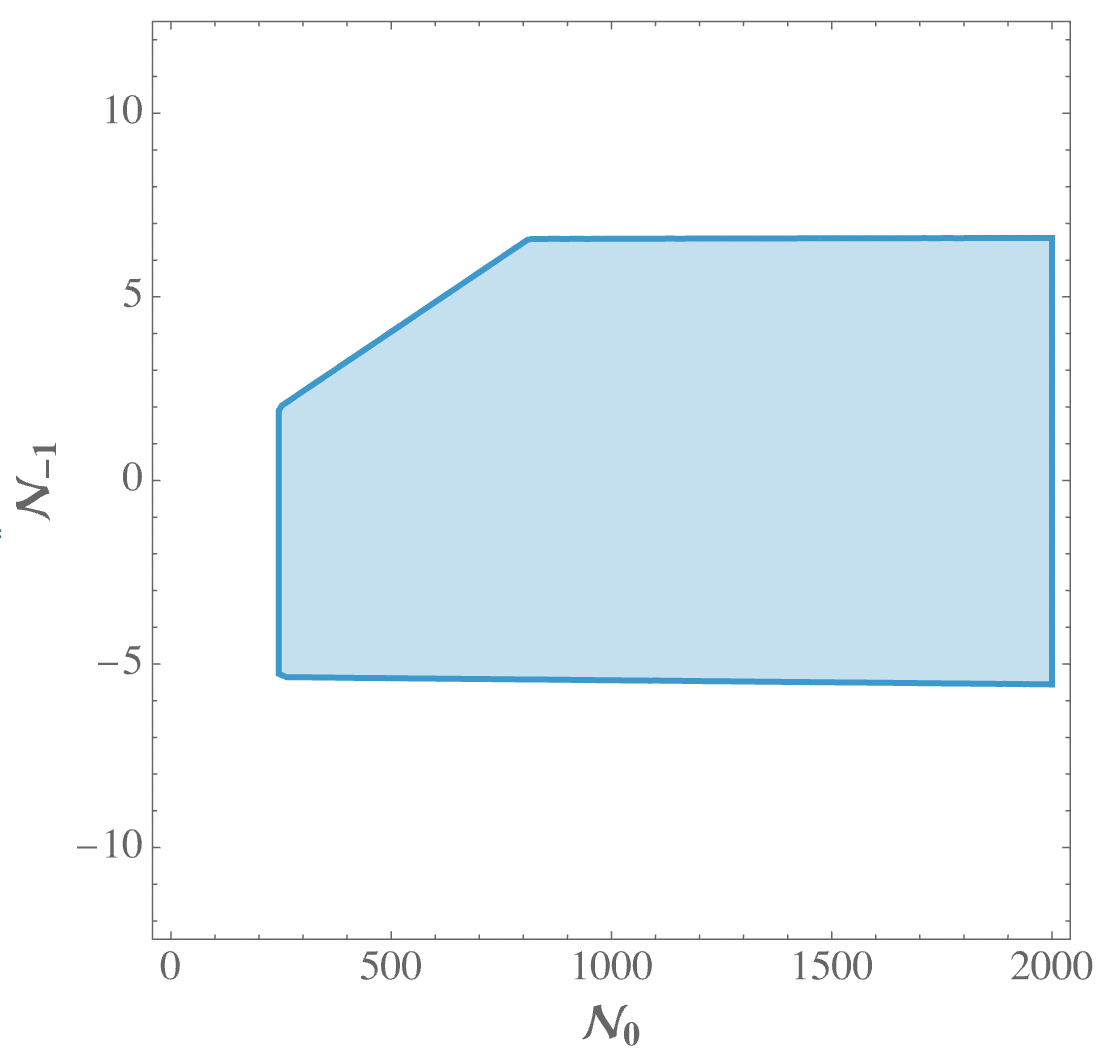}
\caption{Admissible region to ${\cal O}(q^4)$ for the parameters $\cN_0, \cN_{-1}$ in \eref{ell12char}.}
}
\label{consol}
\end{figure}

We immediately read off $c=25$ and $h=\frac14$, and the valence formula then gives $\ell=12$. But when is this admissible? For the identity character to start with unity, we must set $\cN_1=1$. Thereafter the admissible region for $\cN_0,\cN_{-1}$ is whatever makes each coefficient in \eref{ell12char} non-negative. We see that this, in particular, imposes both lower and upper bounds on $\cN_{-1}$. Now at every order in $q$ a new bound emerges, so it may appear that the system is highly overdetermined. 

However we can easily see that there is a non-empty solution space. Taking the intersection of all the bounds to order $q^4$ one finds the space of possible admissible solutions in the blue region in Figure \ref{consol}. To see that some part of this region persists even after we include all orders in $q$, we simply take the rank-2 CFT $(\mE_{8,1})^3 \mA_{1,1}$. This has $c=25$ and $\ell=12$. Its characters are given by \eref{ell12char} with $\cN_0=992, \cN_{-1}=3$, which -- as expected -- lies inside the allowed region plotted in Fig. \ref{consol}.

From these examples we see that the question of which coefficients are negative, and how they grow, is an important input into the procedure of finding admissible characters. In order to systematise the latter procedure, we therefore need to address this question -- to which we now turn.

\section{Asymptotic behaviour and signs of $a_n$}

In this section we consider the simplest limits of the Frobenius recursion relations of sub-section \ref{Frobrecursubsec} in which the behaviour of quasi-characters can be extracted. These give us some confirmation of the properties that were guessed based on examples. In the following section we analyse the more difficult intermediate region between the limits considered here, which will enable us to prove the desired results completely.

\subsection{$n\gg c$}

We start with the asymptotic behaviour for large $n$ and fixed $c$, of the coefficients $a_n$ of different quasi-characters. This can be obtained from the Rademacher formula (see \cite{Chandra:2018pjq} for more details):
\be
a_{j,n}\sim \sum_{i=0}^{1}\sum_{\alpha_i+m<0}
\Big(S_{ji}^{-1}e^{4\pi \sqrt{|\alpha_i+m|(\alpha_j+n)}}+\cdots\Big)a_{i,m}
\label{radem}
\ee
where $S_{ij}$ is the modular (collapsed) $S$-matrix.

We want to find the asymptotic growth in terms of the leading coefficients of the two characters. Hence we choose $m=0$ in \eref{radem}, to get:
\be
a_{j,n}\sim \sum_{i=0}^1\sum_{\alpha_i<0}
\Big(S_{ji}^{-1}e^{4\pi \sqrt{|\alpha_i|(\alpha_j+n)}}+\cdots\Big)a_{i,0}
\label{simpleRad}
\ee

Recall that just after \eref{alphaellzero} we showed that one of the exponents $\alpha_0,\alpha_1$ is positive and the other negative. In fact for $c>0$ we have $\alpha_0<0, \alpha_1>0$ while for $c<0$ we have  $\alpha_0<0, \alpha_1>0$. Thus in the former case, the leading contribution to the RHS of \eref{simpleRad} comes from $i=0$, while in the latter, the leading term comes from $i=1$. We learn that:
\be
\begin{split}
c>0:\qquad~ a_{0,n} &\sim
\exp\left(4\pi\sqrt{\frac{c}{24}\Big(-\frac{c}{24}+n\Big)}\,\right)a_{0,0}\\[2mm]
a_{1,n} &\sim 
\exp\left(4\pi\sqrt{\frac{c}{24}\Big(\frac{c+4}{24}+n\Big)}\,\right)a_{0,0}\\[2mm]
c<0: \qquad a_{0,n} &\sim 
\exp\left(4\pi\sqrt{\left|\frac{c+4}{24}\right|\Big(-\frac{c}{24}+n\Big)}\,\right)a_{1,0}\\[2mm]
a_{1,n} &\sim -
\exp\left(4\pi\sqrt{\left|\frac{c+4}{24}\right|\Big(\frac{c+4}{24}+n\Big)}\,\right)a_{1,0}
\end{split}
\label{asympgrowth.0}
\ee
The last line tells us that asymptotically $a_{1,n}$ is negative if we normalise the non-identity character to start with a positive value.
It follows that for $\ell=0$, all the $c>0$ quasi-characters are of Type I, while all the $c<0$ quasi-characters have a non-identity component of Type II (this terminology was defined in sub-section \ref{qcreview}).  These predictions are confirmed in Table \ref{A1signs.0} and easily verified for the other quasi-character families as well.

\subsection{$c\gg n$}

Here we consider the opposite limit where $c$ is very large compared to $n$ the order in the $q$-expansion. Empirically, as seen in Table \ref{A1signs.0}, the identity character for $c=24M+j>0$ has exactly $M$ negative coefficients, namely $a_1,a_3,\cdots, a_{2M-1}$ while the non-identity character is completely positive. On the other hand for $c=24M+j<0$, the identity character is completely positive while the non-identity has $|M|-1$ positive coefficients namely $a_2,a_4,\cdots a_{2|M|-2}$, always assuming that we have chosen $a_0>0$. Thus in both cases, precisely at $n=2|M|$ there is a crossover from alternating behaviour to a fixed sign, which takes place in the identity character for $M>0$ and in the non-identity character for $M<0$. We now prove part of the above result: namely that for $n\ll c$, the alternating signs indeed arise as seen above. Later we will examine the crossover region. In the following discussion we set $a_0=1$. 

\subsection*{Identity character}

The Frobenius recursion relation for the identity character was written in \eref{frobrecur.l0c0}. In the limit of large $c$ with fixed $n$ the formula simplifies to:
\be
\begin{split}
a_{0,n} &\sim \frac{\sum_{i=1}^n \frac{5c^2}{12}\sigma_3(i)}{\left(-\frac{cn}{12}\right)}a_{0,n-i}\\
&= -\frac{5c}{n}\sum_{i=1}^n \sigma_3(i)a_{0,n-i}
\end{split}
\label{climid.0}
\ee
This implies that each $a_{0,r}$ scales to leading order as $c^r$, so in the above sum one can neglect all but the $i=1$ term. Since $\sigma_3(1)=1$, this leads to:
\be
a_{0,n} \sim \left(-\frac{5c}{n}\right)^n+\cO\left(c^{n-1}\right)
\label{anlargec.l0c0}
\ee
Thus for $c>0$, the sign of $a_{0,n}$ alternates, with negative signs for odd $n$.  This is consistent with the behaviour seen in explicit examples in Table \ref{A1signs.0} (top left quadrant). On the other hand for $c<0$, \eref{anlargec.l0c0}
 is positive for all $n$, again consistent with Table \ref{A1signs.0} (bottom left quadrant).

\subsection*{Non-identity character}

Here we take the large-$c$ limit of \eref{frobrecur.l0c0} to get:
\be
\begin{split}
a_{1,n} &\sim \frac{\sum_{i=1}^n\frac{5c^2}{12}\sigma_3(i)}{\left(\frac{cn}{12}\right)}a_{1,n-i}\\
& \sim \frac{5c}{n}\sum_{i=1}^n \sigma_3(i)a_{1,n-i}
\end{split}
\label{climnonid.0}
\ee
At large $c$ the recursion is dominated as before by the highest power of $c$ so we get:
\be
a_{1,n} \sim \left(\frac{5c}{n}\right)^n+\cO\left(c^{n-1}\right)
\label{anlargec.l0c1}
\ee
This time we see that all coefficients are positive for $c>0$ while they alternate in sign for $c<0$, confirming the remaining quadrants of \ref{A1signs.0} (top right and bottom right). 

Thus the relevant part of our empirical observations is proved. As emphasised from the beginning, the results in this section can only be trusted for $c\gg n$, which was our starting assumption. Hence the region $n\sim c$ is not accessible to the analysis here. To study this region, we introduce a new approach in the following section.

\section{Behaviour for $n\sim |\sfrac{c}{12}|\gg 1$}

Recall that we have parametrised $c$ as $24M+j$ where $1 \le j \le 8$. Thus we are now considering the limit $n\sim |2M| \gg 1$. In explicit examples, we saw that the alternating component of a $\ell=0$ quasi-character exhibits the following features: in some cases there is a crossover from alternating behaviour to a definite sign at $n=2|M|$. The previous asymptotic analyses were carried out for $n\gg c$ or $n \ll c$ and in these limits one obviously cannot see such crossovers. We need to focus instead on the regime $n\sim |c|$. Of course the coefficients can be explicitly computed for any finite value of $n$ and $c$, so we focus on the limit where both $c$ and $n$ are large and of comparable order.  This will permit us to find approximate solutions to the Frobenius recursion in general, and use those to predict the observed behaviour.

\subsection{Recursion relation in the limit}

Recall that in Eqs. (\ref{recurf}), (\ref{fMni.0}) we rewrote the Frobenius recursion by defining the quantity $f_M(n,i)$ and replacing  $c$ by $24M+j$. 
We start by taking the limit of large $M$ where $n\lesssim 2|M|$. For these values of $n$ the second term in the numerator of \eref{fMni.0} dominates over the first as was seen in \eref{climid.0}. But this time we do not neglect $n$ relative to $M$ in the denominator, so (assuming for $n\ne 2|M|$) we get:
\be
\begin{split}
f_M(n,i)~ &{\buildrel |M|\to\infty\over\simeq} ~ \frac{240 M^2}{n(n-2M)}\sigma_3(i)\qquad \hbox{(identity)}\\[2mm]
f_M(n,i)~ &{\buildrel |M|\to\infty\over\simeq} ~ \frac{240 M^2}{n(n+2M)}\sigma_3(i)\qquad \hbox{(non-identity)}
\end{split}
\label{fMnilim1}
\ee
The cases $n=2M, M>0$ for the identity character, and $n=-2M, M<0$ for the non-identity, must be considered separately -- here the expression \eref{fMnilim1} is not valid, and must be replaced by:
\be
\begin{split}
f_M(2M,i) ~ &{\buildrel |M|\to\infty\over\simeq} ~ -\frac{1440 M}{j+2}\sigma_3(i) \qquad \hbox{(identity, $M>0$)}\\
f_M(2|M|,i) ~ &{\buildrel |M|\to\infty\over\simeq} ~ ~~~\frac{1440 |M|}{j+2}\sigma_3(i) \qquad \hbox{(non-identity, $M<0$)}
\end{split}
\label{nonstan}
\ee
Putting everything together, we have:
\be
\begin{split}
f_M(n,i) < 0 \hbox{ for: }& M>0 \hbox{ and } n \leq 2M \qquad \hbox{(identity)} \\
& M<0 \hbox{ and } n< 2|M| \quad \hbox{(non-identity)}\\
f_M(n,i) > 0 \hbox{ for: }& M<0 \quad \hbox{(identity)}\\
& M>0 \quad \hbox{(non-identity)}
\end{split}
\ee
One direct consequence is that the identity character is positive for $M<0$ while the non-identity is positive for $M>0$. Also, both are completely positive for $n>2|M|$. This is because at every step of the recursion, the relevant functions $f_M(n,i)$ are positive. 

Before considering the other cases, let us obtain some bounds.
In \eref{fMnilim1}, for fixed $n$ the coefficient multiplying $\sigma_3(i)$ grows linearly with $|M|$. On the other hand if $n$ is of order $|M|$, say $n=\alpha|M|$ where $\alpha$ is a constant of order 1, then -- for the identity character -- $f_M(n,i)$ tends to a constant:
\be
\begin{split}
f_M(n,i) &\simeq \frac{240}{\alpha(\alpha-2)}\sigma_3(i),\quad M\ge 0\\
&\simeq \frac{240}{\alpha(\alpha+2)}\sigma_3(i),\quad M < 0
\end{split}
\label{fMnilim2}
\ee
The situation is reversed for the non-identity character.

Thus for large $|M|$ and all $n\lesssim 2|M|,~ i\le n$, we have the lower bound:
\be
\begin{split}
|f_M(n,i)| &~>~ \frac{240}{\alpha|\alpha- 2|}, \quad M>0\\
&~>~\frac{240}{\alpha(\alpha+ 2)}, \quad M<0
\end{split}
\label{fbound}
\ee
The lower bound over the full range $\alpha\in(0,2)$ is therefore:
\be
\begin{split}
|f_M(n,i)| &~>~240, \qquad M>0\\
& ~>~30, ~\qquad M<0
\end{split}
\label{fMlb}
\ee

Now we proceed to determine the signs and estimate the magnitudes of $a_n$ for $n=\alpha|M|$. For this, we first write down a formal solution to the recursion relation \eref{recurf} as follows:
\be
\begin{split}
a_n &=\sum_{i_1=1}^n f_M(n,i_1)a_{n-i_1}\\
&=f_M(n,n)a_0+\sum_{i_1=1}^{n-1}\sum_{i_2=1}^{n-i_1} f_M(n,i_1)f_M(n-i_1,i_2)a_{n-i_1-i_2}\\
&=\Big(f_M(n,n)+f_M(n,n-1)f_M(1,1)\Big)a_0\\
&\qquad +\sum_{i_1=1}^{n-2}\sum_{i_2=1}^{n-i_1-1} \sum_{i_3=1}^{n-i_1-i_2} f_M(n,i_1)f_M(n-i_1,i_2)f_M(n-i_1-i_2,i_3) a_{n-i_1-i_2-i_3},\quad \hbox{etc.}
\end{split}
\label{recurexpand}
\ee
Continuing until only $a_0$ appears on the RHS, we see the following structure. The RHS is a sum of $p$ products of $f_M$ factors with $1\le p \le n$. The term of order $p$ has sums over all $i_k, 1\le k \le p$ with the constraint $\sum_{k=1}^p i_k=n$. Every term multiplies $a_0$, which we henceforth set to 1.

We can classify the terms by ordered partitions of $n$. For example the term with partition $(n)$ corresponds to the first term $f_M(n,n)$ in the second line of \eref{recurexpand}, while the term with partition $(1^n)$ corresponds to $i_1=i_2=\cdots i_n=1$ and can be written explicitly as:
\be
\prod_{r=1}^n f_M(r,1)
\ee
Next, the ordered partition $(2.1^{n-2})$ corresponds to $p=n-1$ and we have $i_1=2; i_2,i_3,\cdots, i_{n-1}=1$. The associated term is: 
\be
f_M(n,2)\prod_{r=1}^{n-2} f_M(r,1)
\ee
By summing over all orderings, we can replace ordered by un-ordered partitions. For example the un-ordered partition $(2.1^{n-2})$ corresponds to:
\be
\sum_{k=1}^{n-1} f_M(k+1,2)\prod_{\genfrac{}{}{0pt}{1}{r=1}{r\ne k,k+1}}^n f_M(r,1)
\ee
which can more conveniently be written as:
\be
\left(\prod_{r=1}^n f_M(r,1)\right)\sum_{k=1}^{n-1}\frac{f_M(k+1,2)}{f_M(k,1)f_M(k+1,1)}
\ee

These examples suggest that a useful way to organise all terms in the sum is to first factor out: 
\be
P_M(n)\equiv \prod_{r=1}^n f_M(r,1)
\label{PMdef}
\ee
In the example above, we are left with a ratio $\frac{f_M(k+1,2)}{f_M(k,1)f_M(k+1,1)}$ wherever there is a 2 in the partition replacing two 1s. More generally, we define:
\be
g_M(k,m)\equiv \frac{f_M(k+m-1,m)}{\prod_{r=1}^{m} f_M(k+r-1,1)}
\label{gdef}
\ee
Note that $g_M(k,1)=1$. Intuitively, the combinatorial meaning of each $g_M(k,m)$ is ``start with a chain of $n$ symbols each equal to 1, then at the $k$th position in the chain, delete the next $m$ factors of 1 and replace them by a single factor of $m$''.

Now the subset of terms on the RHS of the recursion relation corresponding to all un-ordered partitions of the form $(m, 1^{n-m})$, for any $m\ge 2$, is (after factoring out $P_M(n)$) given by:
\be
\sum_{m=2}^n\sum_{k=1}^{n-m+1} g_M(k,m)=\sum_{k=1}^{n-1}g_M(k,2)+ \sum_{k=1}^{n-2}g_M(k,3)+\cdots +g_M(1,n)
\label{singleg}
\ee
These are all the terms linear in $g_M$. Quadratic terms in $g_M$ correspond to partitions of the form $(m_1.m_2.1^{n-m_1-m_2})$ with $m_1,m_2\ge 2$. These terms take the form:
\be
\begin{split}
&\sum_{k_1=1}^{n-3}\sum_{k_2=k_1+2}^{n-1}g_M(k_1,2)g_M(k_2,2)+
\sum_{k_1=1}^{n-5}\sum_{k_2=k_1+3}^{n-3}g_M(k_1,3)g_M(k_2,3)\\
&\qquad\quad +
\sum_{k_1=1}^{n-4}\sum_{k_2=k_1+3}^{n-1}g_M(k_1,3)g_M(k_2,2)+
\sum_{k_1=1}^{n-4}\sum_{k_2=k_1+2}^{n-1}g_M(k_1,2)g_M(k_2,3)+\cdots
\\
& = \sum_{m_1m_2\ge 2} \sum_{k_1=1}^{n+1-m_1-m_2}\sum_{k_2=k_1+m_1}^{n+1-m_2}g_M(k_1,m_1)g_M(k_2,m_2)
\end{split}
\label{doubleg}
\ee
The generalisation to higher powers of $g_M$ is straightforward.

In this way we organise the recursion relation \eref{recurf} into contributions of increasing orders in $g_M$. For fixed $n$, the maximum order in $g_M$ is achieved when every term is a $g_M(k,2)$ for some $k$ (because $g_M(k,1)=1$ as noted above). The corresponding order is $\frac{n}{2}$ and corresponds to the partition $2^{\frac{n}{2}}$ (assuming $n$ is even). 

Implicit in our discussion above is the notion that $g_M$ is small and orders in $g_M$ can be used to develop a perturbative expansion. In fact, in the approximation leading to \eref{fMnilim1}, we have:
\be
\begin{split}
g_M(k,m) &\simeq \frac{\sigma_3(m)}{(240M^2)^{m-1}}\prod_{r=1}^{m-1}(k+r-1)(k+r-1-2M)\qquad \hbox{(identity)}\\
&\simeq \frac{\sigma_3(m)}{(240M^2)^{m-1}}\prod_{r=1}^{m-1}(k+r-1)(k+r-1+2M)\qquad \hbox{(non-identity)}
\end{split}
\label{glim1}
\ee
Let us now see what this tells us about the two characters of our solution.

\subsubsection*{Identity character}

Let us now pick the identity character and estimate the value of the product $P_M(n)$ as defined in \eref{PMdef} for large $|M|$. For $M>0$ we use the first line of \eref{fMnilim1}. Because the last argument of each $f_M$ is 1, each of the $f_M$ factors in $P_M$ is proportional to $\sigma_3(1)=1$ and hence that factor can be ignored. Then we get:
\be
\begin{split}
P_M(n) &\simeq\prod_{r=1}^n \frac{240 M^2}{r(r-2M)} \\
&= (-1)^n\frac{(240 M^2)^n(2M-n-1)!}{n!(2M-1)!}\qquad \hbox{ for }n<2M
\end{split}
\label{PMnpos}
\ee

We will be particularly interested in the value at $n\simeq 2|M|$. 
However the answer is apparently divergent at this point, since -- as we already explained -- the approximation \eref{fMnilim1} to $f_M$ is invalid at this value. So the last term in the product \eref{PMdef} that defines $P_M(n)$ has to be replaced by \eref{nonstan}, giving:
\be
P_M(2M)\simeq \frac{1440M}{j+2} \frac{(240 M^2)^{2M-1}}{\big((2M-1)!\big)^2},\quad M>0
\label{PM2Mpos}
\ee
On the other hand, for $M<0$ we get:
\be
\begin{split}
P_M(n) &\simeq\prod_{r=1}^n \frac{240 M^2}{r(r+2|M|)} \\
&= \frac{(240 M^2)^n(2|M|)!}{n!(2|M|+n)!}
\end{split}
\label{PMnneg}
\ee
which is true for any $n\lesssim 2|M|$ and even $n$ larger than $|M|$ as long as it is of the same order.

\subsubsection*{Non-identity character}

For the non-identity case the roles of $M>0$ and $M<0$ are reversed. Noting that the sign of \eref{nonstan} changes as a result, we have in particular:
\be
\begin{split}
P_M(2M) &\simeq \frac{(240 M^2)^{2M}}{(4M)!},\quad M>0\\
P_M(2|M|) &\simeq -\frac{1440|M|}{j+2} \frac{(240 M^2)^{2|M|-1}}{\big((2|M|-1)!\big)^2},\quad M<0
\end{split}
\label{PM2M.nonid}
\ee

Now we can use the recursion to find the sign and estimate the magnitude of the coefficients $a_n$ for $n\simeq 2|M|$.

\subsection{A crude approximation}\label{crude}

\subsubsection*{Identity character}

Consider the identity character, and start with $M>0$. Because $|f_M(n,i)|\ge 240$ for $n<2M$ we can, as an initial approximation, solve the recursion \eref{recurf} by keeping only the $i=1$ term on the RHS at every step. Then we have:
\be
a_{n} \simeq \prod_{r=1}^{n} f_M(r,1) \equiv P_M(n)
\label{ell0est}
\ee
Then for $n<2M$, every factor in the above product is negative. Hence in this approximation the $a_n$ are alternating in sign for all $n<2M$ and have the same sign as $(-1)^n$. This precisely confirms the prediction from explicit calculations. 

It remains to consider $n=2M$ as well as $n>2M$. We saw above that at $n=2M$, the term $n-2M$ in the denominator of \eref{fMni.0} vanishes and therefore we must use \eref{nonstan} instead. Thus we have:
\be
a_{2M}\simeq \left(-\frac{1440 M}{j+2}\right)\prod_{r=1}^{2 M-1} f_M(r,1)
\label{modrelpos.0}
\ee
The terms inside the product are all negative for $\alpha<2$, so the RHS is a product of $2M$ negative numbers and hence positive. Thus we conclude that $a_n$ has the same sign as $(-1)^n$ for all $n$ up to {\em and including} $n=2M$.

Finally, for $n>2M$ (but not too large) we get the expression:
\be
a_{n} \simeq \prod_{r=2M+1}^{n}f_M(r,1)
\left(-\frac{1440 M}{j+2}\right)
\prod_{r=1}^{2M-1} f_M(r,1)
\label{ng2M.0}
\ee
Now all terms in the first product are positive and we already argued that the rest of the expression is positive, so the final answer is positive. We see that once $n$ crosses $2M$ all terms turn positive, in agreement with explicit calculations. Extrapolating to very large $n$, this also matches on correctly to the Rademacher formula \eref{asympgrowth.0}.

For $M<0$ things are even simpler. The above expressions are replaced by the result:
\be
a_{n} \simeq \prod_{r=1}^{n} f_M(r,1)
\ee
for $n \lesssim 2|M|$, and all factors are positive, in perfect agreement with the explicit calculations. However in this case we have $|f_M| > 30$ and this lower bound is almost an order of magnitude smaller than for $M>0$, where it was 240. So we may expect the crude approximation to be less accurate for $M<0$. This will indeed turn out to be the case.

To check the validity of our approximations, let us estimate the growth of both $a_{2|M|}$ and $P_M(2|M|)$ with $|M|$. First let $M>0$. Extrapolating from explicit computations \footnote{The computation was done for the $\mA_1$ family, i.e. $j=1$ in \eref{jvalell.0}, but is expected to give similar answers for all families, since $j\sim\cO(1)$ as $1\leq j <8$.}, we find that the value of $a_{2M}$ up to $M=1800$ (central charge $c=19200$) is consistent with exponential growth:
\be
a_{2M}\sim e^{\gamma M},\quad\gamma\sim 12.13
\label{a2Mgrowth.0}
\ee
Now let us estimate $P_M(2M)$. From \eref{PM2Mpos} we get, for large $M$:
\be
\begin{split}
P_M(2M)&\sim \frac{e^{2M\log 240 M^2}}{e^{4M \ln (2M/e)}}\\
&\sim e^{12.19 M}
\end{split}
\label{PM2Mgrowthpos}
\ee
which is extremely close.

As a check of these results, we can find an upper bound for $|a_{2M}|$ by extrapolating the Rademacher expansion (valid for $n\gg M$) down to $n=2M$ -- though strictly this is not within its range of validity. Then from \eref{asympgrowth.0} we get:
\be
|a_{2M}|\sim e^{4\pi |M|}\sim e^{12.6 M}
\label{Radestimate}
\ee
This applies equally for positive and negative $M$. One can also try to extrapolate upwards from the estimate in \eref{anlargec.l0c0} which was obtained for $M\gg n$ (again going beyond its range of validity), which gives:
\be
|a_{2M}|\sim e^{2M\ln 60}\sim e^{8.2 M},\quad M>0
\ee
providing a comparable lower bound. Both the actual value and our estimate lie squarely within the region bounded by these extrapolated estimates.

Next consider $M<0$. We have already confirmed that $a_n$ is positive in this case. As noted above, the upper and lower bounds coming from the Rademacher expansion (\eref{Radestimate}) and \eref{anlargec.l0c0} are the same as for positive $M$. On the other hand, explicit calculation up to around $M=-1800$ tells us that:
\be
a_{2|M|}\sim e^{\gamma|M|},\quad \gamma \sim 8.09,\quad M<0
\label{a2Mcalc}
\ee
For $P_M(2|M|)$ we use \eref{PMnneg} to find:
\be
\begin{split}
P_M(2|M|) &\sim \frac{e^{2M\ln 240 M^2}}{e^{4M\ln (4M/e)}}\\
&\sim e^{4M(\ln 15.5M-\ln 4M+1)}\\
&\sim e^{9.42M}
\end{split}
\label{PM2Mgrowthneg}
\ee
Comparing Eqs (\ref{a2Mcalc}) and (\ref{PM2Mgrowthneg}) we see that agreement is reasonable, though not as close as for positive $M$. This can likely be attributed to the fact (see the discussion below \eref{fMlb}) that while $f_M > 240$ for positive $M$, for negative $M$ we can only say that $f_M> 30$ so approximations that involve dropping inverse powers of $f_M$ are more risky in that case.

To summarise, for the identity character the crude approximation \eref{ell0est} completely confirms the sign properties that were previously guessed at from examples, and also estimates quite well the growth of the coefficient $a_{2|M|}$ with $M$, with the estimate for positive $M$ being closer than that for negative $M$.

\subsubsection*{Non-identity character}

From \eref{fMnilim1} we see that the estimates for the non-identity character are related to those for the identity by sending $M\to -M$. Hence we find that in the non-identity case, the alternating behaviour of the $a_n, n<2|M|$ as well as the remarkable accuracy of our crude approximation hold true for $M<0$, while the positivity of $a_n$ is verified for $M>0$. One small twist is that for $M<0$, the formula for $a_{2|M|}$ is:
\be
a_{2|M|}\simeq \left(\frac{1440 |M|}{j+2}\right)\prod_{r=1}^{2 |M|-1} f_M(r,1)
\label{modrelneg.0}
\ee
Here we have an odd number of negative terms, so $a_{2|M|}$ is negative. After this point the series stops oscillating and therefore remains negative for $n>2|M|$. This correctly matches onto the prediction from the Rademacher formula valid for $n\gg 2|M|$.

\subsection{A better approximation}

We now examine corrections to \eref{ell0est}, which take the forms in Eqs (\ref{singleg}), (\ref{doubleg}) and so on. These terms are organised by the number of $g_M$ factors in each one. We will try to extract the relevant corrections for large $n\sim 2|M|$. One striking discovery will be that, despite the excellent agreement between Eqs (\ref{a2Mgrowth.0}) and \ref{PM2Mgrowthpos}), the crude approximation of the previous sub-section is somewhat misleading -- there are correction terms of opposite sign that can (and do) overcome the leading terms retained in the crude approximation. However by summing a partial set of corrections we will see that the original predictions are retained.

\subsubsection*{Identity character}

Starting out with the identity character and $M>0$, we define:
\be
\begin{split}
G_M(m) &\equiv \sum_{k=1}^{2M-m+1} g_M(k,m)\\
& \simeq  \frac{\sigma_3(m)}{(240M^2)^{m-1}} \sum_{k=1}^{2M-m+1}\prod_{r=1}^{m-1}(k+r-1)(k+r-1-2M)
\end{split}
\label{GMdef}
\ee
where we used \eref{glim1}. Thus the corrections linear in $g_M$, \eref{singleg}, become:
\be
1+\sum_{m=2}^n G_M(m)
\label{singleG}
\ee
Now it is easily shown (using Mathematica) that:
\be
\begin{split}
\lim_{M\to\infty} \frac{G_M(m+1)}{G_M(m)}&\sim\frac{\sigma _3(m+1) \Gamma
   \left(-m-\frac{1}{2}\right) \Gamma
   (m+1)}{240\, \sigma _3(m) \Gamma
   \left(\frac{1}{2}-m\right) \Gamma
   (m)}\\
   &=-\frac{m\,\sigma_3(m+1)}{240(m+\half)\sigma_3(m)}
\end{split}
\label{GMratio}
\ee
For large $m$ this tends to $-\frac{1}{240}\frac{\sigma_3(m+1)}{\sigma_3(m)}$. Since the generalised divisor function $\sigma_3(m)$ grows as $m^3$, this ratio tends to $-\frac{1}{240}\frac{(m+1)^3}{m^3}\sim -\frac{1}{240}$ for large $m$, and the series $\sum_{m=2}^{2M} G_M(m)$ is convergent. Moreover it is well-approximated by:
\be
\sum_{m=2}^{2M} G_M(m)\simeq G_M(2)\sum_{j=0}^\infty \left(-\frac{1}{240}\right)^j=\frac{240}{241}G_M(2)
\label{wellapprox}
\ee
which shows that the linear corrections can be well-approximated by just $G_M(2)$.

As a check, we compute $\frac{G_M(3)}{G_M(2)}$. For such small values of $m$ one needs  to use the second line of \eref{GMratio} and with this one finds:
\be
\begin{split}
G_M(2)=\sum_{k=1}^{2M-1} g_M(k,2) &\simeq -\frac{M}{20}\\
G_M(3) =\sum_{k=1}^{2M-2} g_M(k,3) &\simeq \frac{7 M}{13500}
\label{G23}
\end{split}
\ee
Thus for $n=2M$, both $G_M(2)$ and $G_M(3)$ grow linearly for large $M$. However an encouraging sign is that $G_M(3)$ is much smaller. Their ratio is approximately:
\be\label{subl}
\frac{G_M(3)}{G_M(2)}\simeq -0.01037
\ee 
The calculated value of the same ratio is in fact found to settle down to $-0.0104$ for sufficiently large $M$ (the error arises from the fact that we neglected the first term in the numerator of \eref{fMni.0} and also neglected some small terms along the way). This is a useful confirmation of our estimates in \eref{fMnilim1}. It disagrees with $-\frac{1}{240}$ by a factor $\sim 2.4$, but that was expected due to $m$ being small. This discrepancy will only affect the first few terms in \eref{singleG} and will not significantly affect \eref{wellapprox}.

\Comment{
Now we use this to estimate the second and third terms in \eref{singleg}:
\be
\begin{split}
\sum_{k=1}^{n-1} g_M(k,2) &\simeq \frac{9}{240 M^2}\sum_{k=1}^{n-1} k(k-2M)\\
& = \frac{1}{160 M^2} n(n-1)  (2n - 6M - 1)
\end{split}
\ee
and:
\be
\begin{split}
\sum_{k=1}^{n-2} g_M(k,3) &\simeq \frac{28}{(240 M^2)^2}\sum_{k=1}^{n-2} k(k-2M)(k+1)(k+1-2M)\\
& = \frac{7}{216000 M^4} (n-2) (n-1) n \left(20
   M^2-15 M n+15 M+3 n^2-6
   n+1\right)
\end{split}
\ee

}

Now let us see what this discussion tells us about the sign and magnitude of the answer. 
We have previously discussed the signs of the functions $f_M$ defined in Eqs (\ref{fMni.0}) from which we may deduce the signs of the  $g_M$ of \eref{gdef}. For the identity character and with positive $M$, we find that the sign of $g_M(k,m)$ is given by $(-1)^{m-1}$, so it is negative for even $m$ and positive for odd $m$. Hence the single-$g_M$ series in \eref{singleg} has terms of alternating signs for $M>0$. Then the magnitude of the corrections becomes important since they can affect the overall sign. On the other hand for negative $M$, all the $g_M$ functions are positive and they cannot change the sign of the answer.

Now from \eref{GMdef} we find that for $M>20$ the magnitude of $G_M(2)$ becomes greater than 1 at $n\simeq 2M$, showing that it can potentially change the sign of the full sum. Moreover as \eref{wellapprox} shows, adding the remaining linear terms $G_M(m)$ makes little difference. Thus it is necessary to consider terms involving higher than linear powers of the $g_M$.

As a start we consider only those terms involving multiple powers of $g_M(k,2)$, such as the first term in \eref{doubleg}, which we recall is:
\be
\sum_{k_1=1}^{n-3}\sum_{k_2=k_1+2}^{n-1}g_M(k_1,2)g_M(k_2,2)
\ee
Now we add the corresponding terms with cubic and higher powers of $g_M(k,2)$. We will find that all such terms can be approximately summed. To start with, since we are taking $n$ large, we replace the upper limits of both sums in the above expression by $n$. Now since $k_2>k_1$, the sum covers half the space $k_1,k_2\in \{1,n\}$. So we  approximate the quadratic term above by:
\be
\half \left(\sum_{k=1}^{n} g_M(k,2)\right)^2 
\ee
Similarly the contribution of terms with $p$ powers of $g_M(k,2)$ becomes at large $n$:
\be
\frac{1}{p!}\left(\sum_{k=1}^{n} g_M(k,2)\right)^p
\ee
As a result, the series exponentiates. Of course there are corrections that we expect (but have not rigorously proven) to be small. So we get:
\be
a_{n}\simeq P_M(n)\,
e^{\sum_{k=2}^{n} g_M(k,2)}
\ee
Using the first line of \eref{G23} we thus have:
\be
\begin{split}
a_{2M} &\simeq P_M(2M)\, e^{-\frac{M}{20}}
\end{split}
\label{a2Mcorrpos}
\ee
This exponentiation immediately solves our sign problem: the sign of $a_{2M}$, and more generally $a_n$ for $n<2M$ as well as $n$ slightly above $2M$, will be the same as that of $P_M(n)$. This cements the agreement with explicit calculations. 

Now for the magnitude of $a_{2M}$, the exponential factor leads to a small correction to the ``crude'' approximation based only on \eref{PM2Mgrowthpos}. Using that equation in \eref{a2Mcorrpos} we now get:
\be
a_{2M}\simeq e^{(12.19-0.05)M}=e^{12.14 M}
\label{a2Mcloser}
\ee
This brings the answer even closer to the result extrapolated by calculating up to $M=1800$, which was $12.13$. This is very satisfying \footnote{In fact, for $M=2000$, we get the actual value as $a_{2M} = e^{12.1358}$ and the better approximated value as $a_{2M} = e^{12.1385}$ where the percentage error in the $\log$ values is about $0.02\%$}.

The above discussion was for $M>0$. Repeating the same steps for $M<0$, we get:
\be
\begin{split}
G_M(m) &\equiv \sum_{k=1}^{2|M|-m+1} g_M(k,m)\\
& \simeq  \frac{\sigma_3(m)}{(240M^2)^{m-1}} \sum_{k=1}^{2|M|-m+1}\prod_{r=1}^{m-1}(k+r-1)(k+r-1+2|M|)
\end{split}
\label{GMdef2}
\ee
This is an apparently minor change with respect to \eref{GMdef}. However from explicit calculation we find that the ratio $G_M(m+1)/G_M(m)$ is significantly larger for negative than for positive $M$. Thus for example $G_M(101)/G_M(100)\simeq 0.0039$ for $M>0$ while the same ratio is $0.029$ for $M<0$ \footnote{In fact in the two cases the ratios seem to stabilise around $0.004\sim \frac{1}{240}$ and $0.03\sim \frac{1}{30}$ respectively, consistent with the bounds on $f_M$ in \eref{fMlb}.}. Yet it is still small enough that $\sum_{m=2}^{2|M|} G_M(m)$ can be approximated by $G_M(2)$. The rest of the argument goes through as before and we find that:
\be
a_{2|M|} \simeq P_M(2|M|)\,e^{\frac{M}{4}}
\ee
This time the correction in the exponential is positive, and the growth in \eref{PM2Mgrowthneg} is modified from $e^{9.42 M}$ to $e^{9.67 M}$. This actually worsens the comparison with the explicit calculation for which the growth is found to be roughly $e^{8.09 M}$. Fortunately this is a case where the sign is not in question to start with, since all terms and corrections are positive. 

\subsubsection*{Non-identity character}

Like the leading estimate, the corrections in this case are found essentially by exchanging $M\leftrightarrow -M$.

\section{Inductive approach to signs of $a_n$}\label{induct}

In this section, we provide an inductive argument to show that the sign of $a_n$ is the same as sign of $(-1)^n$. Using this result in appendix \ref{appB} we will show that the exact expression for $a_n$ as in \eref{recurexpand} can be written as an exponential function of $M$ and a real positive parameter $\beta$ which we define in that appendix. This would imply that for $1\leq n\leq 2M$, $a_n$ as given in \eref{recurexpand} nicely exponentiates!

Here we will be able to prove the desired alternation of signs without making any approximations, overcoming some of the limitations of the previous section. While this proves our key result regarding signs, it does not provide estimates for the growth of coefficients similar to \eref{a2Mcloser}. However, it enables us to find the following growth patterns of the coefficients:
\begin{equation}
\begin{split}
    &a_n \geq R \, a_{n-1}, \quad \text{with } R>1 \, \, \text{and } \, 1\leq n \leq 2|M|
\end{split}    
\end{equation}
which we dub as the {\it super-geometric} growth to contrast it to the geometric growth which happens for $R=1$. Note that from the Rademacher formula in \eref{asympgrowth.0}, we have a geometric growth of coefficients in the regime of $n\gg c$ (where Rademacher approximation is valid). Thus, the necessary condition for Rademacher behaviour to kick in is $\left|\frac{a_{n+1}}{a_n}\right|\simeq 1$. From numerics we have: for the identity character with $M=50 \, (c\sim 1200)$ with $j=1$ and $n=50M^2 = 125000$, the ratio being $\left|\frac{a_{n+1}}{a_n}\right|\simeq 1.134$. We further observed that this ratio keeps on decreasing and asymptoting towards unity as we increase $M$ in the above numerics.

From the above discussion, we see that if we consider the crude approximation as in Sec. \ref{crude} and compute the ratio $\left|\frac{a_{n+1}}{a_n}\right|$ for the identity character with $M > 0$, then it doesn't hold beyond $n\sim 16 M$ since around this point, we get from \eref{PMdef}: $\frac{P_M(n+1)}{P_M(n)}=f_M(n+1,1)\simeq 1$ for large $M$. Also, for any scaling of the kind $n = c_1 M^2$ with $c_1 >1$, $\frac{P_M(n+1)}{P_M(n)}=f_M(n+1,1)\to 0$ for large $M$. Thus, we actually have very different behaviours in the regimes: $n\ll |M|$, $n\sim \cO(|M|)$ and $n \gg M$ like $n\sim \cO(|M|^2)$.

\subsection{Bounds on the divisor function}

To carry out the inductive proof we first need to find bounds on various functions that arise in it. Here we bound $\sigma_1(i)$ and $\sigma_k(i), k\ge 2$.
\begin{align}\label{sig1_0}
    \sigma_1(i) = \sum_{d|i} d = \sum_{e|i}\frac{i}{e} = i\sum_{e|i}\frac{1}{e} < i\sum_{k=1}^i\frac{1}{k}
\end{align}
with $e\equiv\frac{i}{d}$. Now $\sum_{k=1}^i\frac{1}{k} \leq 1 + \int_1^i \frac{dx}{x} \leq 1+ \log(i)$. Thus, we get,
\begin{align}\label{sig1_1}
    \sigma_1(i)\leq i(1+\log(i))
\end{align}
For $\sigma_k(i), k\ge 2$, we have:
\begin{align}\label{sigk}
    \sigma_k(i) = \sum_{d|i} d^k = \sum_{e|i} \left(\frac{i}{e}\right)^k  < i^k\sum_{n=1}^\infty \frac{1}{n^k} = i^k\zeta(k)
\end{align}    
Since $\zeta(2)\simeq 1.6$ and $\zeta(k)\to 1$ monotonically as $k\to\infty$, we have $\zeta(k)<2$ for all integral $k\ge 2$. Combining this with the obvious lower bound $\sigma_k(i)\geq i^k$, which is saturated at $i=1$, we get:
\be
i^k < \sigma_k(i)< \zeta(k)i^k < 2i^k,\qquad k\ge 2
\label{sigkbounds}
\ee

\subsection{Bounds on the functions $f_M(n,i)$}\label{l0_id_Mg1}
In this section, we shall find bounds on the numerator and denominator of $f_M(n,i)$ as in \eqref{fMni.0}. This, in turn, will bound $f_M(n,i)$ itself. For simplicity, we will outline the procedure for $M>0$ and for the identity character only, though some equations below will hold generically for $M<0$ and for the non-identity character too. 

Using the bounds of the previous section, let us find out where the numerator $N_{n,i}$ and denominator $D_n$ change signs, where:
\be
\label{Ni_Dn}
f_M(n,i)\equiv\frac{N_{n,i}}{D_n}
\ee
where $f_M(n,i)$ was defined in \eref{fMni.0}. For the identity character, we get:
\begin{align}\label{NDdef}
N_{n,i} &\equiv \frac{5}{12}(24M+j)(24M+j+4)\sigma_3(i) - 4\left(n-i-M-\frac{j}{24}\right)\sigma_1(i) \equiv A\sigma_3(i) - B\sigma_1(i) \nonumber\\
D_n &\equiv n\left(n-2M-\frac{j+2}{12}\right)
\end{align}
We see that $N_{n,i}$ vanishes at:
\begin{align}\label{gen_i}
    n = \frac{A\sigma_3(i)}{4\sigma_1(i)} + i + M + \frac{j}{24},
\end{align}
and goes from positive to negative as $n$ increases through this value. 
Let us consider $i=1$ above, which gives:
\begin{align}\label{i_1}
    n_1\equiv\frac{A}{4} + 1 + M + \frac{j}{24} = 60 M^2 + (5j + 11)M + 1 +  \left\lceil\frac{5j^2 + 22j}{48}\right\rceil.
\end{align}
For general $i\leq n$ we find from Eqs.(\ref{sig1_1}) and (\ref{sigkbounds}) that $\frac{\sigma_3(i)}{\sigma_1(i)}\geq \frac{i^2}{1+\log(i)}\geq 1$, which saturates at $i=1$. In particular, $\frac{\sigma_3(i)}{\sigma_1(i)}$ is an increasing function of $i$ and hence for $i>1$, no $N_{n,i}<0$ before $N_{n,1}$ \footnote{Till the above equation, that is, \eref{i_1}, the expressions are valid for $M<0$ case also.}.

Now let us take $M>0$. We find bounds on $N_{n,i}$. Expanding $A$ in \eref{NDdef}, we see that:
\be
A =240M^2+20(j+2)M+\frac{5}{12}j(j+4)
\ee
from which we find (with $M>0$):
\be
240M^2+60M < A < 240M^2+240M\label{Abounds}
\ee
Similarly for $B$ we find:
\be
-4(M+1) < B < 4M 
\ee

Hence the bounds on $N_{n,i}$ are:
\be
(240M^2+60M) i^3-4iM(1+\log i)< N_{n,i} < 2(240M^2+240M)i^3+4(M+1)i, 
\label{Nnibounds}
\ee
The above is a general bound for all $n\geq 1$. Later on we will find stronger bounds for specific ranges of $n$.

Next we consider the denominator $D_n$. It changes sign from negative to positive precisely at $n=2M+1$. Hence:
\begin{align}
    &f_M(n,i)<0, \ \ \ \ \text{for }n\leq 2M, \nonumber\\
    &f_M(n,i)>0, \ \ \ \ \text{for }2M< n<60M^2.
\end{align}
Above will be precisely the two regimes of $n$ that we will be focusing on during the inductive proof for the identity character \footnote{For simplicity, we will take the upper bound in the second regime of $n$ to be $50M^2$ instead.}.

Now to bound $|D_n|$, we note that in the range $1\leq n\leq 2M$ it attains its maximum value at $n_*=M+\frac{j+2}{24}$, with:
\be
|D_{n_*}|\simeq M^2+\frac{j+2}{12}M
\label{max_Dn}
\ee
Its minimum value is attained at $n=2M$ where we get $|D_n|=\frac{M(j+2)}{12}$. 
This gives the bounds
\be
M<|D_n|< M^2+M
\label{Dbounds}
\ee

Putting everything together and recalling \eref{Ni_Dn} we find:
\be
|f_M(n,i)| ~\ge~ \left(60 + \frac{180 M}{M+1}\right)\,i^3-\frac{4i}{M+1}(1+\log i)
\label{modfMni.lb}
\ee

\subsubsection*{$\mathbf{1} \leq \bm{n} \leq \mathbf{2}\bm{M}$}
Here, we will restrict for the first range: $1\leq n\leq 2M$. Let us put $n=1$ in \eref{modfMni.lb} to get:
\begin{align}\label{bdd_fmn1}
    |f_M(n,1)| = \frac{N_{n,1}}{|D_n|} > \frac{8(30M+7)}{M+1}  \geq 148,
\end{align}

Next, using the upper bound on $N_{n,i}$, for $i\ge 2$ and the lower bound on $N_{n,1}$ in \eref{Nnibounds}, we find that:
\be
\label{rat_bdd}
\begin{split}
\frac{|f_M(n,i)|}{|f_M(n,1)|} = \frac{N_{n,i}}{N_{n,1}} &\leq 2i^3+ \frac{10}{6M+1}i^3 + \frac{M+1}{10M(6M+1)}i\\
& < 4i^3
\end{split}
\ee
Here the last inequality follows because the maximum value of the RHS in the second-last inequality is reached at $M=1$, at which point the RHS becomes: $\frac{24}{7}i^3 + \frac{1}{35}i < 4 i^3$.

Eqs. \eqref{bdd_fmn1} and \eqref{rat_bdd} are the kind of inequalities that we will use in the inductive proof later.

\subsubsection*{$\bm{2M} < \bm{n} \leq \bm{50M^2}$}
Now let us try to find a bound for $f_M(n,i)$ in the second range of interest, $2M<n\leq 50M^2$. Following the same procedure used to derive \eref{Nnibounds}, we get:
\begin{equation}\label{60M2_l0}
\begin{split}
    &N_{n,i} >\left(240M^2 + 60 M + 2\right)i^3 - 4i(1+\log i)\left(50 M^2 - M - 1\right), \\
    &N_{n,i} < (291M^2 + 243M + 49)i^3 + 4(M+1)i
\end{split}    
\end{equation}
These bounds are stronger than \eref{Nnibounds} because we have restricted the range of $n$. This leads to:
\begin{align}
    \frac{f_M(n,i)}{f_M(n,1)} &= \frac{N_{n,i}}{N_{n,1}}\\
    &\leq \frac{(291M^2 + 243M + 49)i^3 + 4(M+1)i}{40M^2 + 64 M +6} \\
    &< \frac{291 M^2 + 247 M + 53}{40M^2 + 64 M +6}i^3, \nonumber\\
    &<8 i^3,  
\end{align}
To obtain the last inequality, we used the fact that $\frac{291 M^2 + 247 M + 53}{40M^2 + 64 M +6}$ is an increasing function of $M$ that tends to 7.28 as $M\to\infty$.

\subsection{Identity character}\label{Mg0_id}

\subsubsection*{$\bm{M} > \bm{0}$ case}

In this section, we will inductively show the following: in the range $1\leq n \leq 2M$, $a_n<0$ for odd $n$, $a_n>0$ for even $n$; and in the range $2M < n < n_1\simeq 60M^2 + \mathcal{O}(M)$, $a_n>0$. For us it will be sufficient to consider the range $2M<n\leq 50M^2$.

For the region $n>50 M^2$, we get $\frac{M}{n}\sim \frac{c}{n} \sim \frac{\cO(M)}{\cO(M^2)}\to 0$, for large $M$, implying that we are in the regime of $n\gg c$. Hence we can use Rademacher asymptotics as in \eref{asympgrowth.0} to determine the signs of $a_n$ for $n>50M^2$. Together with the inductive proof, this tells us that $a_n>0$ for all $n>2M$.

\subsubsection*{$\bm{1}\leq \bm{n}\leq \bm{2M}$}

Let us define $b_n\equiv (-1)^n a_{n}$. In the given range, $f_M(n,i)<0$ and the recursion relation \eref{recurf} becomes:
\begin{align}\label{b_n}
    b_n = \sum_{i=1}^n (-1)^{n-(n-i)} f_M(n,i) b_{n-i} = \sum_{i=1}^n (-1)^{i+1} |f_M(n,i)| b_{n-i}
\end{align}
Moreover from $a_0=1, a_1<0$ we easily see that $b_0, b_1 > 0$. Now we would like to show that $b_n>0$ for all $n\le 2M$. This in turn will imply that $a_n$ is non-zero and has the same sign as $(-1)^n$ for $n\leq 2M$.

From \eref{b_n} we find:
\be
\begin{split}
b_n &= |f_M(n,1)|b_{n-1} - \sum_{i=2}^n (-1)^{i}|f_M(n,i)| b_{n-i}\\
& > |f_M(n,1)|b_{n-1} - \sum_{i=2}^n |f_M(n,i)| b_{n-i}
\end{split}
\label{expand_bn1}
\ee
So if we can prove that for any given value of $n$,
\begin{align}\label{topr}
    |f_M(n,1)|b_{n-1} > \sum_{i=2}^n |f_M(n,i)| b_{n-i},
\end{align}
then we will have proved $b_n>0$ (in the given range $1\le n\le 2M$).

In order to show this, we will first show that there exists a number $R$ satisfying $1<R\leq 148$ -- independent of $n$ -- such that $b_n \geq R b_{n-1}$ for all $n\le 2M$. This in turn will imply that all the $b_n$ are positive, which is our goal. The result is already true for $b_1$, since $b_1 = |f_M(n,1)|b_0 \geq 148\, b_0$ by \eref{bdd_fmn1}. Next, fix $n$ and assume the result to be true for all $k<n$. Then we will prove it holds for $k=n$. 

The induction hypothesis, when iterated, tells us that:
\begin{align}\label{chain_bk}
    b_{n-i} \leq  \frac{b_{n-1}}{R^{i-1}},\qquad i\ge 2
\end{align}
This in turn implies:
\begin{align}
    \sum_{i=2}^n|f_M(n,i)|b_{n-i} \leq b_{n-1}\sum_{i=2}^n\frac{|f_M(n,i)|}{R^{i-1}}
\label{fbbound}
\end{align}
From Eqs (\ref{expand_bn1}) and (\ref{fbbound}) we get:
\be
\begin{split}\label{bk-1}
    b_n & > |f_M(n,1)|b_{n-1} - \sum_{i=2}^{n}|f_M(n,i)|b_{n-i}\\
    &\geq 148\, b_{n-1} - b_{n-1}\sum_{i=2}^n\frac{|f_M(n,i)|}{R^{i-1}}\\
    &> 148\, b_{n-1}\left(1-4\sum_{i=2}^\infty\frac{i^3}{R^{i-1}}\right) \\
    &>\alpha(R)\,b_{n-1}
\end{split}
\ee
where:
\be
\alpha(R)=148\,\frac{R^4-36R^3+26R^2-20R+5}{(R-1)^4}
\label{alphaR}
\ee
In the third line of \eref{bk-1} we used \eref{rat_bdd} and then extended the sum to $\infty$ which only decreases the RHS and therefore preserves the inequality. In the last step we evaluated the infinite sum.

Now if we can prove $\alpha(R)\gtrsim R$ then we are done. This in turn will hold if there is a real solution to $\alpha(R)=R$ with $\alpha(R)$ given by \eref{alphaR}. Numerically we find there are three real solutions, $\simeq 0.33, 53.35, 97.93$, with $\alpha(R)>R$ between the latter two. Thus for $R$ anywhere in the range $54<R<97$, the desired inequality holds.

Thus, we have proved that $b_n>Rb_{n-1}$ for all $n\leq 2M$. This now implies that $b_n>0$ and are monotonic. This in turn shows that $a_n$ are alternating up to $n=2M$.

\subsubsection*{$\bm{2M} < \bm{n} \leq \bm{50M^2}$}
Now we make an inductive argument for $n$ in the range $2M<n\leq 50M^2$ and show that $a_n>0$ for all $n$. For this, let us split the recursion relation into two parts:
\begin{align}\label{g2M+1}
    a_n = \sum_{i=1}^{n-2M} f_M(n,i)a_{n-i} + \sum_{i=n-2M+1}^n f_M(n,i)a_{n-i}
\end{align}
With this, first let us show that $a_{2M+1}>0$ which will serve as the base case for the next inductive argument.

From \eref{g2M+1}, we have:
\begin{equation}\label{2M+1_0}
\begin{split}
    a_{2M+1} &= f_M(2M+1,1)a_{2M} + \sum_{i=2}^{2M+1}f_M(2M+1,i)a_{2M+1-i} \\
    &\geq f_M(2M+1,1)a_{2M}-\sum_{i=2}^{2M+1}f_M(2M+1,i)|a_{2M+1-i}|
\end{split}    
\end{equation}
Now we use the fact that $|a_n|\geq R |a_{n-1}|$ for $1\leq n\leq 2M$, to argue that $|a_{2M+1-i}|\leq \frac{|a_{2M}|}{R^{i-1}}$, with the same range of values $R\in(54,97)$ that we obtained above. Using this and $a_{2M}>0$, we find from \eref{2M+1_0} that:
\begin{align}\label{2M+1_001}
    a_{2M+1} \geq f_M(2M+1,1)a_{2M} - a_{2M}\sum_{i=2}^{2M+1}\frac{f_M(2M+1,i)}{R^{i-1}}
\end{align}

Now using \eref{60M2_l0} in \eref{2M+1_001} we get:
\begin{align}
    a_{2M+1} &\geq f_M(2M+1,1)a_{2M}\left(1-\sum_{i=2}^\infty \frac{8\, i^3}{R^{i-1}}\right) \geq f_M(2M+1,1)a_{2M} > 0
\end{align}
since: $1-\sum_{i=2}^\infty\frac{8 i^3}{R^{i-1}}>0$ for $R\in(68,97)$. This range is contained in the original one $R\in (54,97)$. So the base case is done: $a_{2M+1}>0$. Now our induction hypothesis is $a_{n}>0$ for $2M+1 < n < 50M^2$, and we want to show that $a_{50M^2}>0$.

Consider $n=50M^2$ and let us split the recursion relation in \eref{recurf} into two parts as follows,
\begin{equation}\label{eqn009}
    a_n = \sum_{i=1}^{n-2M-1}f_M(n,i)a_{n-i} + \sum_{i=n-2M}^n f_M(n,i)a_{n-i},    
\end{equation}
where the first term on the RHS is positive by our induction hypothesis and in the second term we have alternating signs for $a_{n-i}$ since $n-i\in\{0,\ldots,2M\}$. Now, a simple re-arrangement of terms allows us to write the second term in the RHS of \eref{eqn009} as:
\begin{align}\label{split_2}
    \sum_{i=n-2M}^n f_M(n,i)a_{n-i} &= \sum_{k=0}^{M-1} f_M(n,n-2M+2k)|a_{2M-2k}| \nonumber\\
    &\qquad -  \sum_{k=0}^{M-1} f_M(n,n-2M+2k+1)|a_{2M-2k-1}| + f_M(n,n)a_0 \nonumber\\
    &\geq \sum_{k=0}^{M-1}|a_{2M-2k}|\left[f_M(n,n-2M+2k) - \frac{f_M(n,n-2M+2k+1)}{R}\right] \nonumber\\
    &\qquad + f_M(n,n)a_0 \nonumber\\
    &\geq \sum_{k=0}^{M-1}|a_{2M-2k}|f_M(n,n-2M+2k+1)\left[\frac{f_M(n,n-2M+2k)}{f_M(n,n-2M+2k+1)} - \frac{1}{R}\right] \nonumber\\
    &\qquad + f_M(n,n)a_0,
\end{align}
where, in the second last step, we have used the fact that: $|a_{2M-2k-1}|\leq \frac{|a_{2M-2k}|}{R}$ for all $0\leq k\leq M-1$ and $R\in(68,97)$ -- which was proven above. Now if we can show that the term in the square bracket above is positive then that would imply that the whole RHS of \eref{eqn009} is positive leading to $a_n>0$ for $n=50M^2$. So, let us show this next.

Note that for all $i\equiv n-2M+2k$ with $0\leq k\leq M-1$ we have $50M^2 \geq i\geq 50M^2-2M > 48 M^2$. From the definitions in \eref{NDdef} and using similar procedure as outlined in Sec. \ref{l0_id_Mg1} we get the following bounds:
\begin{equation}
\begin{split}
    N_{n,i} &\geq i^3\left(A-\frac{4M}{i^2}(1+\log i)\right), \\
    N_{n,i+1} &\leq A\zeta(3)(i+1)^3 + 4M(i+1), \\
    \frac{N_{n,i}}{N_{n,i+1}}&\geq \frac{A-\frac{4M}{i^2}(1+\log i)}{A\zeta(3)(1+i^{-1})^3 + 4M\frac{i+1}{i^3}}\\ &> 0.304,
\end{split}
\end{equation}
where to get to the last inequality note that: $\frac{A-\frac{4M}{i^2}(1+\log i)}{A\zeta(3)(1+i^{-1})^3 + 4M\frac{i+1}{i^3}}$ is an increasing function of $i$, $M$ and $j$ with its least value being $i_{\text{min}}=50 M^2 - 2M$, $M=1$ and $j=1$ which yeilds $0.782$. Thus,
\begin{align}
    \frac{N_{n,i}}{N_{n,i+1}} = \frac{f_M(n,i)}{f_M(n,i+1)} = \frac{f_M(n,n-2M+2k)}{f_M(n,n-2M+2k+1)}>0.782
\end{align}
for all $0\leq k\leq M-1$. Hence, for the RHS of \eref{split_2} to be positive we need to satisfy $0.782-\frac{1}{R}>0$ implying $R\gtrsim 1.23$ which is true since we had picked some $R\in(54,97)$. Hence, we are done.

\subsection*{$\bm{M}<\bm{0}$ case}
Here let us consider $|M|>0$ in the first equation of \eref{fMni.0}, to get:
\begin{align}\label{Ml0_id}
    f_M(n,i) &\equiv\frac{N_{n,i}}{D_n}\equiv \frac{-4 \big(n-i+|M|-\frac{j}{24}\big)\sigma_1(i)+\frac{5(j-24|M|)(j+4-24|M|)}{12}\sigma_3(i)}{n\Big(n+2|M|-\frac{j+2}{12}\Big)}.
\end{align}
From above we get that the denominator $D_n>0$ for all $n>0$. For the numerator, we have its zero at:
\begin{align}\label{n_i_Ml0}
    n_i = i + \frac{j}{24} + \left(\left[\frac{5j}{12} + \frac{5j^2}{48} - 10|M| - 5j|M| + 60|M|^2\right]\frac{\sigma_3(i)}{\sigma_1(i)} - |M|\right),
\end{align}
and $N_{n,i}$ changes its sign from positive to negative as $n$ increases through $n_i$. Now the term in parentheses in \eref{n_i_Ml0} is positive and is also an increasing function of $|M|$. Additionally, we know that $\frac{\sigma_3(i)}{\sigma_1(i)}$ is an increasing function of $i$ for $i\geq 1$. This further implies that $n_i>0$ and is also an increasing function of $i$ and $|M|$. Thus, like before, the zero of $N_{n,1}$, denoted by $n_1$, satisfies $n_1<n_i$ for all $2\leq i\leq n$. So, $N_{n,i}>0$ for all $1\leq i\leq n < n_1$. 

Next let us determine $n_1$. From \eref{Ml0_id} we get,
\begin{equation}\label{lower_n1}
\begin{split}
    n_1 = 60|M|^2 + 1 + \frac{11j}{24} + \frac{5j^2}{48} - |M|(5j + 11) > 50|M|^2
\end{split}    
\end{equation}
since $|M|\geq 1$. Hence, $f_M(n,i)>0$ for all $1\leq n\leq 50|M|^2$. Now we know that $a_0=1$ and hence by induction on $n$ we get from the recursion relation \eref{recurf} that $a_n > 0$ since it is a sum of positive terms. Beyond $n=50|M|^2$, we are in the regime: $n\gg M$ and we can invoke the Rademacher expression as in \eref{asympgrowth.0} to claim that $a_n>0$.
    
Now let us prove the super-geometric growth of the coefficients $a_n$ for $1\leq n\leq 2|M|$. Firstly, note that, using the same procedure to find bounds as outlined in Sec. \ref{l0_id_Mg1}:
\begin{equation}\label{lbdd1}
\begin{split}
    &N_{n,i}\geq 50|M|^2 i^3 - 4|M|i(1+\log i) \geq N_{n,1}\geq 50|M|^2 - 4|M|, \\
    &D_n \leq 2|M|\left(4|M|-\frac{j+2}{12}\right) \leq 8|M|^2,
\end{split}
\end{equation}
which implies,
\begin{align}
    f_M(n,i)\geq f_M(n,1) \geq \frac{25}{4}-\frac{1}{2|M|} \geq 5
\end{align}
Now pick any $R\in(1,5)$. Using this in the recursion \eref{recurf} we get:
\begin{equation}
\begin{split}
    a_n &= f_M(n,1)a_{n-i} + \ldots + f_M(n,n)a_0, \\
    &\geq R(a_{n-1} + \ldots + a_0) \geq R a_{n-1}
\end{split}
\end{equation}
where we have used that $a_{k}>0$ for all $0\leq k\leq n-2$. Hence, proved.

\subsection{Non-identity character}

\subsubsection*{$\bm{M} < \bm{0}$ case}

Now we will inductively show the following: in the range $1\leq n < 2|M|$, $a_n<0$ for odd $n$, $a_n>0$ for even $n$; and in the range $2|M| \leq n \lesssim n_1\simeq 60|M|^2-49|M|$, $a_n<0$. For us it will be sufficient, to take the range of $n$ in the second part to be $2|M|\leq n \leq 15|M|^2$.

For the region where $n>15 |M|^2$, note that $\frac{|M|}{n}\sim \frac{|c|}{n}\to 0$ since $c\sim |M|$ and $n\sim |M|^2$ implying, that we are in the regime of $n \gg |c|$ and hence we can presumably use Rademacher asymptotics as in \eref{asympgrowth.0} to determine the signs of $a_n$ for $n>15 |M|^2$ which further leads to the conclusion that $a_n<0$.

\subsubsection*{$\bm{1}\leq \bm{n}<\bm{2|M|}$}
Using \eref{fMni.0} for the non-identity character we get:
\begin{align}\label{fmni00}
    f_M(n,i) &\equiv \frac{N_{n,i}}{D_n} \equiv \frac{-4 \big(n-i-|M|+\frac{j+4}{24}\big)\sigma_1(i)+\frac{5(j-24|M|)(j+4-24|M|)}{12}\sigma_3(i)}{n(n-2|M|+\frac{j+2}{12})}
\end{align}
With this, using the same procedure to find bounds as outlined in Sec. \ref{l0_id_Mg1}, we get these useful bounds:
\begin{equation}\label{useful_nonid_mg1}
\begin{split}
    (240 |M|^2 - 200|M| + 40)i^3 -4|M|i(1+\log i) < &N_{n,i} < (240|M|^2 - 60|M| + 2)\zeta(3)i^3 + 4|M|i \nonumber\\
    \frac{|M|}{3}-\frac{1}{6} < &|D_n| < |M|^2 - \frac{|M|}{6}
\end{split}
\end{equation}
which implies $f_M(n,i)<0$ for all $1\leq n < 2|M|$. This further leads to the following bounds on $f_M(n,i)$:
\begin{equation}\label{useful_nonid_mg2}
\begin{split}
    |f_M(n,1)| &> 91 \\
    \frac{|f_M(n,i)|}{|f_M(n,1)|}\equiv\frac{N_{n,i}}{N_{n,1}} &< \zeta(3)i^3 + \frac{144\zeta(3)|M|i^3 + 4|M|i - 38\zeta(3)i^3}{240|M|^2 - 204|M| + 40} \\
    &< 2.6 i^3
\end{split}    
\end{equation}
where to prove the last inequality above we need to show:
\begin{align*}
    &2.6(240|M|^2 - 204|M|+40) - \zeta(3)(240|M|^2 - 60|M|+2)i^2 > 4|M|,
\end{align*}
Since $i\geq 1$, it is sufficient to prove the above inequality for $i=1$. For this we define:
\begin{align*}
    Q(|M|) \equiv 2.6(240|M|^2 - 204|M|+40) - \zeta(3)(240|M|^2 - 60|M|+2) - 4|M|.
\end{align*}
Note that, $Q(|M|)$ is an increasing function of $|M|$ and $Q(2)\simeq 520 >0$. For $i=1$, we have $Q(1)<0$. However, for $i=1$, we also have $N_1/N_1 = 1 < 2.6$. Hence, we are done.

Now, as before, let us define $b_n\equiv (-1)^n a_{n}$. In the given range, $f_M(n,i)<0$ and the recursion relation \eref{recurf} becomes:
\begin{align}\label{b_ng1}
    b_n = \sum_{i=1}^n (-1)^{n-(n-i)} f_M(n,i) b_{n-i} = \sum_{i=1}^n (-1)^{i+1} |f_M(n,i)| b_{n-i}
\end{align}
Moreover from $a_0=1, a_1<0$ we easily see that $b_0, b_1 > 0$. Now we would like to show that $b_n>0$ for all $n < 2|M|$. This in turn would imply that $a_n$ is non-zero and has the same sign as $(-1)^n$ for $n < 2|M|$.

From \eref{b_ng1} we see that:
\be
\begin{split}
b_n &= |f_M(n,1)|b_{n-1} - \sum_{i=2}^n (-1)^{i}|f_M(n,i)| b_{n-i}\\
& > |f_M(n,1)|b_{n-1} - \sum_{i=2}^n |f_M(n,i)| b_{n-i}
\end{split}
\label{expand_bn1g1}
\ee
So if we can prove that for any given value of $n$,
\begin{align}\label{toprg1}
    |f_M(n,1)|b_{n-1} > \sum_{i=2}^n |f_M(n,i)| b_{n-i},
\end{align}
then we will have proved $b_n>0$ (in the given range $1\le n< 2|M|$).

Now using Eqs. \eqref{useful_nonid_mg1} and \eqref{useful_nonid_mg2} we will show that there exists a number $R$ satisfying $1<R\leq 91$ -- independent of $n$ -- such that $b_n \geq R b_{n-1}$ for all $n < 2|M|$. This in turn will imply that all the $b_n$ are positive, which is our goal. The result is already true for $b_1$, since $b_1 = |f_M(n,1)|b_0 \geq 91\, b_0$ by \eref{useful_nonid_mg2}. 

Next, assume the result to be true for all $k<n$. Then we will prove it holds for $k=n$. The induction hypothesis, when iterated, tells us that:
\begin{align}\label{chain_bkg1}
    b_{n-i} \leq  \frac{b_{n-1}}{R^{i-1}},\qquad i\ge 2
\end{align}
This in turn implies:
\begin{align}
    \sum_{i=2}^n|f_M(n,i)|b_{n-i} \leq b_{n-1}\sum_{i=2}^n\frac{|f_M(n,i)|}{R^{i-1}}
\end{align}
From Eqs (\ref{expand_bn1g1}) and (\ref{chain_bkg1}) we get:
\be
\begin{split}\label{bk-1g1}
    b_n & > |f_M(n,1)|b_{n-1} - \sum_{i=2}^{n}|f_M(n,i)|b_{n-i}\\
    &\geq 91\, b_{n-1} - b_{n-1}\sum_{i=2}^n\frac{|f_M(n,i)|}{R^{i-1}}\\
    &> 91\, b_{n-1}\left(1-2.6\sum_{i=2}^\infty\frac{i^3}{R^{i-1}}\right) \\
    &=\alpha(R)\,b_{n-1}
\end{split}
\ee
where:
\be
\alpha(R)=91\,\left(1-2.6\sum_{i=2}^\infty\frac{i^3}{R^{i-1}}\right)
\label{alphaRg1}
\ee
In the second-last step of \eref{bk-1g1} we used Eqs. \eqref{useful_nonid_mg1} and 
\eqref{useful_nonid_mg2} and then extended the sum to $\infty$ which only decreases the RHS and therefore preserves the inequality. In the last step we evaluated the infinite sum.

Now if we can prove $\alpha(R)\gtrsim R$ then we are done. This in turn can be done if there is a real solution to $\alpha(R)=R$ with $\alpha(R)$ given by \eref{alphaRg1}. Numerically we find there are three real solutions, $\simeq 0.33, 41.64, 52.59$, with $\alpha(R)>R$ between the latter two. Thus for $R$ anywhere in the range $42<R<52$, the desired inequality holds.

Thus we have proved that $b_n>Rb_{n-1}$ for all $n< 2|M|$ which implies 
that $b_n>0$ and are monotonic. This in turn shows that $a_n$ are alternating up to $n=2|M|-1$. 

Next, we see why this result fails above $n=2|M|-1$. This is because $f_M(n,i)>0$ for $n=2|M|$ and hence in \eref{b_ng1}, the leading term becomes negative and hence $b_{2M}<0$ implying $a_{2M}<0$.

\subsubsection*{$\bm{2|M|}\leq \bm{n} < \bm{15 |M|^2}$}

From \eref{fmni00}, we have that $N_{n,1}$ changes sign from positive to negative as $n$ increases through $n_1$ given by:
\begin{equation}\label{n1_Mlll0}
\begin{split}
    n_1 &= 60|M|^2 - (5j+9)|M| + 1 + \left\lceil\frac{j(j+4)}{16}\right\rceil < 60|M|^2 - 49|M| + 7 \\
    &> 15|M|^2
\end{split}
\end{equation}
As before, after some algebra, we see that no $N_{n,i}$ turns negative before $N_{n,1}$. Now let us first show that $a_{2|M|}<0$ which will serve as the base case for our next inductive argument.

Now define $b_{2|M|}\equiv-a_{2|M|}$. From the alternating pattern of $a_n$ above in the range $1\leq n < 2|M|$ and the recursion in \eref{recurf} we get,
\begin{equation}\label{b2M_Ml0}
\begin{split}
    b_{2|M|} &= f_M(2|M|,1)|a_{2|M|-1}| - \sum_{i=2}^{2|M|}(-1)^{i-1}f_M(2|M|,i)a_{2|M|-i} \\
    &\geq f_M(2|M|,1)|a_{2|M|-1}| - \sum_{i=2}^{2|M|}(-1)f_M(2|M|,i)a_{2|M|-i}
\end{split}
\end{equation}
Now let us use the super-geometric growth property -- which we have shown earlier -- to get: $|a_{2|M|-i}|\leq \frac{|a_{2|M|-1}|}{R^{i-1}}$ with $R\in(42,52)$. Using this above we get,
\begin{equation}\label{b2M_nn1}
\begin{split}
    b_{2|M|} &\geq f_M(2|M|,1)|a_{2|M|-1}| - a_{2|M|-1}\sum_{i=2}^{2|M|}(-1)\frac{f_M(2|M|,i)}{R^{i-1}} \\
    &\geq f_M(2|M|,1)|a_{2|M|-1}| - a_{2|M|-1}\sum_{i=2}^{\infty}(-1)\frac{f_M(2|M|,i)}{R^{i-1}}
\end{split}
\end{equation}
Now following the procedure outlined in Sec.\ref{Mg0_id} to get \eref{60M2_l0}, we get,
\begin{align}\label{ratio_bdd12}
    \frac{f_M(n,i)}{f_M{n,1}}<5i^3
\end{align}
for $2|M|\leq n\leq 15|M|^2$ and $1\leq i\leq n$. Using this in \eref{b2M_nn1} we get:
\begin{align}
    b_{2|M|}&\geq f_M(2|M|,1)|a_{2|M|-1}|\left(1-\sum_{i=2}^\infty\frac{5i^3}{R^{i-1}}\right)\nonumber\\
    &\geq f_M(2|M|,1)|a_{2|M|-1}|\nonumber\\
    &> 0,
\end{align}
which holds for any $R\in(44,52)$. This range is contained in the original one $R\in(42,52)$. So, we get: $a_{2|M|}<0$. Hence, we are done proving the base case. Let us define for all $2|M|\leq n\leq 15|M|^2$, $b_n\equiv -a_n$.

Let us form our inductive hypothesis as: $b_n>0$ for all $2|M|+1\leq n < 15|M|^2$. Using this, we want to next prove that $b_n>0$ for $n=15|M|^2$. For this $n$ we get, from the recursion relation \eref{recurf}
\begin{align}\label{bmn_Ml0}
    b_n = -\sum_{i=1}^{n-2|M|}f_M(n,i)a_{n-i} - \sum_{i=n-2|M|+1}^n f_M(n,i)a_{n-i}
\end{align}
where $a_{n-i}$s in the first term of RHS above are negative by our inductive hypothesis and these coefficients in the second term are alternating since for the second term: $n-i\in\{0,\ldots, 2|M|-1\}$. Also, recall that for this regime we have the super-geometric growth of the coefficients. Using this fact in \eref{bmn_Ml0} and after some algebra we get,
\begin{equation}\label{bmn_099}
\begin{split}
    b_n &= \sum_{i=1}^{n-2|M|}f_M(n,i)|a_{n-i}| \\
    &\qquad + \sum_{k=0}^{|M|-1}f_M(n,n-2|M|+2k+1)|a_{2|M|-2k-1}| - f_M(n,n-2|M|+2k+2)|a_{2|M|-2k-2}| \\
    &\geq \sum_{i=1}^{n-2|M|}f_M(n,i)|a_{n-i}| \\
    &\qquad + \sum_{k=0}^{|M|-1}f_M(n,n-2|M|+2k+2)|a_{2|M|-1-2k}|\left[\frac{f_M(n,n-2|M|+2k+1)}{f_M(n,n-2|M|+2k+2)} - \frac{1}{R}\right]
\end{split}    
\end{equation}
Note that, for $i\equiv n-2|M|+2k+1$ with $0\leq k\leq |M|-1$, we have:
\begin{equation}\label{ineq0099}
\begin{split}
    \frac{f_M(n,n-2|M|+2k+1)}{f_M(n,n-2|M|+2k+2)} = \frac{N_{n,i}}{N_{n,i+1}} &\geq \frac{A - \frac{4(|M|+1)(1+\log i)}{i^2}}{A\zeta(3)(1+i^{-1})^3 + 4|M|\frac{i+1}{i^3}} \\
    &> 0.676
\end{split}
\end{equation}
where $A\equiv\frac{5}{12}(j-24|M|)(j+4-24|M|)$. We got to the last inequality above by noting that the lower bound above is increasing in $i$, $M$ and $j$. So, its minimum value occurs at $i_{\text{min}} = 15|M|^2 - 2|M| + 1$, $|M|=1$ and $j=1$ which yields $0.676$. So, from \eref{bmn_099} and \eref{ineq0099} we see that $b_n>0$ if,
\begin{equation}
\begin{split}
    &\frac{N_{n,i}}{N_{n,i+1}} - \frac{1}{R}>0, \\
    \text{or,}& \, \, R > 1.48
\end{split}
\end{equation}
which holds for chosen $R\in(44,52)$. Hence, we have shown $b_n>0$, implying $a_n<0$ for all $2|M|\leq n\leq 15|M^2|$.

\subsubsection*{$\bm{M} > \bm{0}$ case}
From the second equation of \eref{fMni.0}, we get:
\begin{align}\label{Ml0_id1}
    f_M(n,i) &\equiv\frac{N_{n,i}}{D_n}\equiv \frac{-4 \big(n-i+M+\frac{j+4}{24}\big)\sigma_1(i)+\frac{5(j+24M)(j+4+24M)}{12}\sigma_3(i)}{n\Big(n+2M+\frac{j+2}{12}\Big)}.
\end{align}
From above we get that the denominator $D_n>0$ for all $n>0$. For the numerator, we have its zero at:
\begin{align}\label{n_i_Ml01}
    n_i = i - \frac{j+4}{24} + \left(\left[\frac{5j}{12} + \frac{5j^2}{48} + 10M + 5jM + 60|M|^2\right]\frac{\sigma_3(i)}{\sigma_1(i)} - |M|\right),
\end{align}
and $N_{n,i}$ changes its sign from positive to negative as $n$ increases through $n_i$. Now the term in parentheses in \eref{n_i_Ml01} is positive and is also an increasing function of $M$. Additionally, we know that $\frac{\sigma_3(i)}{\sigma_1(i)}$ is an increasing function of $i$ for $i\geq 1$. This further implies that $n_i>0$ and is also an increasing function of $i$ and $M$. Thus, like before, the zero of $N_{n,1}$, denoted by $n_1$, satisfies $n_1<n_i$ for all $2\leq i\leq n$. So, $N_{n,i}>0$ for all $1\leq i\leq n < n_1$. 

Next let us determine $n_1$. From \eref{Ml0_id1} we get,
\begin{equation}\label{lower_n11}
\begin{split}
    n_1 = 60M^2 + 9M + 5jM + \frac{3j}{8} + \frac{5j^2}{48} + \frac{5}{6} > 50 M^2
\end{split}    
\end{equation}
since $M\geq 1$. Hence, $f_M(n,i)>0$ for all $1\leq n\leq 50M^2$. Now we know that $a_0=1$ and hence by induction on $n$ we get from the recursion relation \eref{recurf} that $a_n > 0$ since it is a sum of positive terms. Like before, beyond $n=50M^2$, we are in the regime: $n\gg M$ and we can invoke the Rademacher expression as in \eref{asympgrowth.0} to claim that $a_n>0$.

Now let us prove the super-geometric growth of the coefficients $a_n$ for $1\leq n\leq 2M$. Firstly, note that, using the same procedure to find bounds as outlined in Sec. \ref{l0_id_Mg1}, we get:
\begin{equation}\label{lbdd11}
\begin{split}
    &N_{n,i}\geq 50M^2 i^3 - 12Mi(1+\log i) \geq N_{n,1}\geq 50M^2 - 12M, \\
    &D_n \leq 2M\left(4M+\frac{j+2}{12}\right) \leq 2M (4M+1),
\end{split}
\end{equation}
which implies,
\begin{align}
    f_M(n,i)\geq f_M(n,1) \geq \frac{25M-6}{4M+1} \geq 3
\end{align}
Now pick any $R\in(1,3)$. Using this in the recursion \eref{recurf} we get:
\begin{equation}
\begin{split}
    a_n &= f_M(n,1)a_{n-i} + \ldots + f_M(n,n)a_0, \\
    &\geq R(a_{n-1} + \ldots + a_0) \geq R a_{n-1}
\end{split}
\end{equation}
where we have used that $a_{k}>0$ for all $0\leq k\leq n-2$. Hence, the result is proved.

\section{Conclusions}

We have proved that all quasi-characters of vanishing Wronskian index have alternating signs following the pattern in Table \ref{A1signs.0}. Two types of proofs were given: one that involves an approximation where the central charge is taken large along with $n$, and the other without any approximation, using induction. The approximation allows us to estimate the growth of the coefficient $a_{2|M|}$, which is the one where the alternating sign stabilises to a positive or negative value. This is a region outside the conventional limit in which the Cardy/Rademacher formulas can be applied. The estimate is particularly accurate for the identity character at $M>0$, as seen in \eref{a2Mcloser}.

For apparently technical reasons, the case of Wronskian index $\ell=2$ seems more difficult to study than $\ell=0$ \footnote{The admissible ones, first identified as CFTs in \cite{Gaberdiel:2016zke} and studied further in \cite{Mukhi:2017ugw}, are also considerably more complicated than the MMS solutions at $\ell=0$.}. The alternating signs of quasi-characters follow a slightly different pattern -- in examples, one finds alternation until $n=|M|$, then the reverse alternation (i.e. with one extra negative sign) until $n=2|M|$, followed by stability. We hope to report on this in upcoming work \cite{upDM}. With this, we will have a complete understanding of the quasi-characters that generate all rational CFTs regardless of Wronskian index. 

The regime where the alternating signs stabilise, $n\sim 2|M|\sim \frac{|c|}{12}$, seems to have the following significance. Suppose we take one particular quasi-character and attempt to make a non-chiral ``partition function'' $\sum_{i=0}^1 \chi_i \bchi_i$ from it (this is not admissible and does not correspond to a CFT, but considerations of modular invariance do not know this). This partition function contains a term:
\be
a_n{\bar a}_0\, q^{-\frac{c}{24}+n}{\bar q}^{-\frac{c}{24}}
\ee
corresponding to a state of $(h, {\bar h})=(n,0)$ and hence $\Delta=n$. Now for $n\sim 2M, M>0$ we see that the overall power $E=h+{\bar h}-\frac{c}{12}=n-2M$ tends to zero (in other words, it remains small and finite as $M\to\infty$). In the context of irrational CFT this has been described (see e.g. \cite{Hartman:2014oaa}, \cite{Benjamin:2018kre}) as the boundary between ``light'' and ``medium'' states. It is intriguing that the ``crossover'' takes place at this boundary, perhaps one can find an intuitive reason for this.

Pushing further on the above connections to the irrational case, we note that $c\text{ fixed}$ and $E\to\infty$ regime of \cite{Hartman:2014oaa} (see Eq. (1.3) therein) which is the regime where Cardy growth holds, is equivalent -- under the identifications considered in the above paragraph -- to the regime of $M\text{ fixed}$ and $n\to\infty$ -- which is the regime where Rademacher growth is valid.

Additonally, it is interesting to note that, for $1\leq n \lesssim \frac{|c|}{12}$, we get a super-geometric growth indicating a {\t sparseness} in the distribution of Fourier coefficients $a_n$ in this regime. On the other hand, when $n\gg c$ (where Rademacher kicks in) we have a geometric growth, implying a {\it denseness} of the Fourier coefficients $a_n$. This behaviour could potentially shed some light on the sparseness/denseness of states in genuine rational CFTs (see \cite{Benjamin:2018kre}).

One reason for all of the above connections to the irrational case could simply be modularity. To test this, one can compare the behaviour of these $\rm SL(2,\mathbb{Z})$ quasi-characters to the quasi-characters which were obtained in \cite{Das:2022bxm} as solutions of MLDEs to more generic groups like the Fricke groups of prime level -- of which $\rm SL(2,\mathbb{Z})$ is a sub-group.

Conventional modular forms do not, as far as we know, exhibit the property that the sign of coefficients  alternates up to a point where it stabilises, which makes quasi-characters interesting modular objects. But beyond their mathematical interest they are of practical importance because they are a basis for admissible characters that populate the space of solutions to the holomorphic modular bootstrap. Knowledge of signs and growths is crucial in mapping out this space, since in the sum \eref{qcsum.0}, specific negative contributions to the coefficient of $q^n$ have to be overcome by positive contributions from other terms. The general structure of this space of solutions is currently under investigation.

Generalisation to more than two characters is an open problem. In fact, quasi-characters for the three-character case were studied in \cite{Mukhi:2020gnj}, where infinite families of examples were provided (see also \cite{Govindarajan:2025rgh, Govindarajan:2025jlq, Govindarajan:2026frs, Govindarajan:2026cgv} for recent works in this direction). A complete classification should not be too difficult, and it can then be subjected to the same sort of analysis that was carried out here for the case of two characters. The space of three-character admissible VVMF is very different from that of the two-character case -- already at $\ell=0$ there are infinitely many examples \cite{Mathur:1988gt} as against just 7 examples with two characters. Thus, the picture could be quite different in the two cases.

A different approach to finding admissible characters for $\ell\ge 6$, based on the analysis of MLDE with movable poles, was studied recently in \cite{Das:2023qns}. The results of that paper were compared with those of \cite{Chandra:2018pjq} and complete agreement was found in the domain of comparison. It will be interesting if the quasi-character approach can illuminate the role of accessory parameters, and their associated variety, that were discussed in \cite{Das:2023qns}.

\section*{Acknowledgements}

The work of A.D. is supported by the STFC Consolidated Grant ST/T000600/1 -- ``Particle Theory at the Higgs Centre''. He gratefully acknowledges King's College London and the organisers of ICFT 2025: UK Meetings on Integrable and Conformal Field Theory -- especially Gerard Watts, where preliminary results of this work were presented. He also thanks Jishu Das, Faisal Karimi, Anatoly Konechny, Alexander Radcliffe, Jagannath Santara, Naveen Balaji Umasankar and Gerard Watts for their insightful discussions on RCFTs and modular forms. S.M. is supported by the Raja Ramanna Chair of the Department of Atomic Energy, Government of India. He would also like to thank Yuji Tachikawa and the Institute for the Mathematics and Physics of the Universe (IPMU), University of Tokyo, for their warm and generous hospitality during the period that this work was being completed.

\appendix

\section{Explicit computations of low-lying coefficients}

\label{explicitlow}

Here we tabulate the first few coefficients for the $\mA_{1}$ series of quasi-characters for $0\leq n\leq 8$ and $-3\le M\le 3$, as illustrative examples.

\begin{table}[h!]
\begin{adjustwidth}{-35pt}{0pt}
\centering
{\scriptsize
\begin{tabular}{|c|p{7.6cm}|p{8.4cm}|}
\hline
\Tstrut $M$ & Identity character & Non-identity character \\[1mm]
\hline
\Tstrut 
0 & 
$1 + 3 q + 4q^2 + 7q^3 + 13q^4 + 19q^5
+ 29q^6 + 43q^7 + 62q^8 + \cO(q^9) $
&
$1 + q + 3q^2 + 4q^3 + 7q^4 + 10q^5 + 17q^6 + 23q^7 + 35q^8 + \cO(q^9)$ \\[1mm]
\hline
\hline
\Tstrut
1 & 
$1 - 245q + 142640q^2 + 18615395q^3 
+ 837384535q^4 + 21412578125q^5 + 379389640345q^6 + 5165089068645q^7 + 57498950829715q^8 + \cO(q^9) $
& 
$1105 + 101065q + 3838295q^2 + 88358360q^3 + 1454696521q^4 + 18742858160q^5 + 199800669415q^6 + 1829051591175q^7 + 14763322626790q^8 + \cO(q^9) $ \\[8mm]
\hline
\Tstrut
2 & 
$13 - 4361q + 1024492q^2 - 284433485q^3 + 296843797565q^4 + 84306237909803q^5 + 8867059968079425q^6 + 534104386666020723q^7
+ 21861967373053966060q^8 + \cO(q^9) $ &
$10778285 + 2203389405q + 196852441583q^2 + 10638927572924q^3+ 401976188060447q^4 + 11547256546262990q^5 + 266410265073401785q^6 
+ 5130076445059897723q^7 + 84818281019706292425q^8 + \cO(q^9) $ \\[12mm]
\hline
\Tstrut
3 & $119 - 53363q + 14459256q^2 - 3364790387q^3 + 842188593869q^4 - 303881533638137q^5 + 461207383305660887q^6 + 203501875932273013375q^7 + 34505840401212977669887q^8 + \cO(q^9)$ & $72679207559 + 23350075063007q + 3383269741642441q^2 + 300871659609546368q^3 + 18776600055902020059q^4 + 888536031807642974868q^5 + 33555109821077459224757q^6 + 1048799975659078891924625q^7 + 27882387156891739840510632q^8 + \cO(q^9)$ \\[16mm]
\hline
\hline
\Tstrut
$-1$ & $209 + 13547q + 360088q^2 + 5912219q^3 + 70940547q^4 + 679136057q^5 + 5469901089q^6 + 38382994913q^7 + 240478794445q^8 + \cO(q^9)$ & $1 - 247q - 86241q^2 - 4182736q^3 - 96220123q^4 - 1444316844q^5 - 16273561845q^6 - 148598558649q^7 - 1153094272636q^8 - \cO(q^9)$ \\[8mm]
\hline
\Tstrut
$-2$ & $1747563 + 306927249q + 23306923812q^2 + 1067344426501q^3 + 34209156782563q^4 + 836213984046341q^5 + 16482880547977319q^6 + 272348444677096285q^7 + 3880481233093420304q^8 + \cO(q^9)$ & $11 - 3397q + 796489q^2 - 286457228q^3 - 233881376239q^4 - 31524787414118q^5 - 2067774861752617q^6 - 85699330454705235q^7 - 2550099348343571677q^8 - \cO(q^9)$ \\[12mm]
\hline
\Tstrut
$-3$ & $2228934627 + 650601251681q + 85092063968768q^2 + 6806495569359481q^3 + 381489267486322253q^4 + 16209645539099609007q^5 + 550012236476163613755q^6 + 15464514647907602118167q^7 + 370388802746111712875333q^8 + \cO(q^9)$ & $19 - 7973q + 2069429q^2 - 480847208q^3 + 130883258691q^4 - 66753637072728q^5 - 86276737034684531q^6 - 19552263390203208491q^7 - 2220250925482531884330q^8 - \cO(q^9)$ \\[16mm]
\hline
\end{tabular}}
\caption{First few quasi-characters for the $\mA_1$ series}
\label{A1series.0.qexp}
\end{adjustwidth}
\end{table}

Table \ref{A1series.0.qexp} presents the $q$-expansion of $\mA_1$ quasi-characters in a few cases, and Table \ref{A1signs.0} summarises the behaviour of the signs of the coefficients. In the latter table we specify these signs wherever they are alternating, while ``Type I'' and ``Type II'' refer to whether the coefficients are asymptotically positive or negative respectively (we always take the leading coefficient $a_0$ to be positive). Thus, for example ``II $(a_2,a_4>0)$'' means the series has all negative coefficients except for $a_0, a_2, a_4$ which are all positive.

We see that the following pattern holds: the non-identity character for $M>0$ and the identity character for $M<0$ have all positive coefficients. In the identity character for $M>0$, all the $a_n$ for odd $n\leq 2M-1$ are negative, while the remaining coefficients are all positive. And in the non-identity character for $M<0$, the $a_n$ for even $n\leq 2|M|-2$ are positive, while the remaining coefficients are all negative. We have verified that identical patterns hold for the first few examples of all the remaining quasi-character series under consideration. 

\begin{table}[h!]
\centering
{\scriptsize
\begin{tabular}{|c|c|c||c|c|}
\hline
\Tstrut $M$ & $c$ & Type & $-c-4$ & Type \\[1mm]
\hline
\Tstrut 
0 & $j$ & I & $-5$ & I\\[1mm]
\hline
\hline
\Tstrut
1 & $24+j$ & I ($a_1<0$) & $-28-j$ & I\\
2 & $48+j$ & I ($a_1,a_3<0$) & $-52-j$ & I \\
3 & $72+j$ & I ($a_1,a_3,a_5<0$) & $-76-j$ & I\\
4 & $96+j$ & I ($a_1,a_3,a_5,a_7<0$) & $-100-j$ & I\\
5 & $120+j$ & I ($a_1,a_3,a_5,a_7,a_9<0$) & $-124-j$ & I\\[1mm]
\hline
\hline
\Tstrut
$-1$ & $-24+j$  & I & $20-j$ & II \\
$-2$ & $-48+j$ &  I & $44-j$ & II ($a_2>0$) \\
$-3$ & $-72+j$ &  I & $68-j$ & II ($a_2,a_4>0$) \\
$-4$ & $-96+j$ & I & $92-j$ & II ($a_2,a_4,a_6 >0$)  \\
$-5$ & $-120+j$  & I &
$116-j$ & II ($a_2,a_4,a_6,a_8 >0$) \\[1mm]
\hline
\end{tabular}}
\caption{The sign behaviour of each series}
\label{A1signs.0}
\end{table}

\section{Comparison to Ref. \cite{Chandra:2018pjq}}\label{AppA}

In the classification of \cite{Chandra:2018pjq}, there are four quasi-character series with $\ell=0$:
\be
\begin{split}
\hbox{Lee-Yang series:}\qquad & c=\frac{2(6n+1)}{5},~h=
\frac{n+1}{5},\quad 
n\ne 4~\hbox{mod }5\\[2mm]
\mA_1\hbox{ series:}\qquad & c=6n+1,~h=\frac{2n+1}{4}\\
\mA_2\hbox{ series:}\qquad & c=4n+2,~h=\frac{n+1}{3},\quad n\ne 2~\hbox{mod }3\\
\mD_4\hbox{ series:}\qquad & c=12n+4,~h=n+\half
\end{split}
\label{quasiellzero}
\ee
These values satisfy the valence formula:
\be
c+2=12h
\label{valence.0}
\ee

Notice that the excluded values of $n$ above are those for which $h$ is integral. In these cases one can verify that the Frobenius method gives indeterminate answers for one of the solutions (the one corresponding to the ``identity'' character) which results in a logarithmic solution.

For $\ell=0$, the exponents of the two characters are:
\be
\begin{split}
\alpha_0 &= -\frac{c}{24}\\
\alpha_1 &= -\frac{c}{24}+h=\frac{c+4}{24}
\end{split}
\label{alphaellzero}
\ee
These are interchanged by sending $c\to -c-4$.  In turn, this means we send $n\to -n-2$ (Lee-Yang and $\mA_2$), or $n\to -n-1 ~(\mA_1 \hbox{ and }\mD_1)$. This means in each case it is sufficient to scan one of the two Frobenius solutions over the entire range of $n$ (positive and negative). 

Notice that there are no quasi-characters in this range $-4<c<0$. It then follows from \eref{alphaellzero} that precisely one of the two exponents must be positive and the other negative. We now re-organise the above quasi-characters in a slightly different way from \cite{Chandra:2018pjq} that makes it easier to understand the construction of admissible characters. To do this, let us define the following four series at $\ell=0$:
\be
\begin{split}
\hbox{Lee-Yang series:}\qquad & c=\frac{2(6n+1)}{5},~h=
\frac{n+1}{5},\quad 
n = 0,1,2,3~\hbox{mod }10\\[2mm]
\mA_1\hbox{ series:}\qquad & c=6n+1,~h=\frac{2n+1}{4},\quad n=0,1~\hbox{mod }4\\
\mA_2\hbox{ series:}\qquad & c=4n+2,~h=\frac{n+1}{3},\quad n=0,1~\hbox{mod }6\\
\mD_4\hbox{ series:}\qquad & c=12n+4,~h=n+\half,\quad n~\hbox{even}
\end{split}
\label{quasiellzerorev}
\ee
We have restricted the range of $n$ to cover exactly half the cases in \eref{quasiellzero}. However we now take $n\in \IZ$, so it runs over all positive and negative integers. Then the remaining half of the solution set, that we excluded by so restricting $p$, arises at negative values of $p$ but with the role of the identity and non-identity solution reversed. For example, in the $A_1$ series we have excluded $p=2$ which would have $c=13$, however we now allow $p=-3$ which corresponds to $c=-17$. This is exactly $-c-4$ for $c=13$, which means the $c=13$ solution has re-appeared with an exchange of the two components. Notice that when $c$ is negative, $h$ is also negative and therefore the exponent of the identity character is {\em less singular} than that of the non-identity character.

Now let us split the Lee-Yang series into four series, one for each value of $n$ mod 10. Similarly we split the $\mA_1,\mA_2$ series into two each. Thus we end up with 9 distinct series that have:
\be
c=24M+j, \quad h = 2M+\frac{j+2}{5}
\ee
where $j\in \{\frac25, 1, 2, \frac{14}{5}, 4, \frac{26}{5}, 6,7, \frac{38}{5}\}$. The last step is to notice, as first explained in \cite{Mathur:1988gt}, that for $j=\frac25, \frac{38}{5}$ one of the fusion rules following from the Verlinde formula \cite{Verlinde:1988sn} is negative. Thus these cases cannot correspond to a conventional CFT, and indeed have been categorised as Intermediate Vertex Algebras \cite{Kawasetsu:2014}. It turns out that the case with $c=\frac25$ can be recast as a non-unitary CFT by exchanging the roles of the two characters, and in this form it is fully consistent and corresponds to the famous Lee-Yang CFT \footnote{This explains the name given to this family in \cite{Chandra:2018pjq}, though perhaps a more appropriate name would be ``Fibonacci'' family in view of the corresponding MTC.} that is one of the BPZ minimal models. Hence it remains possible that combining quasi-characters with $c=24M+\frac25$ can lead to non-unitary CFTs. But since non-unitary CFTs are virtually impossible to classify, we restricted our attention to the unitary case by dropping both $j=\frac25$ and $j=\frac{38}{5}$. This leaves the 7 families of quasi-characters listed in \eref{jvalell.0}.

\section{Exponential expression for exact $a_n$}\label{appB}
In this appendix, we demonstrate that for the range $1 \leq n \leq 2M$, the Fourier coefficients $a_n$, whose exact expression is given in equation \eref{recurexpand}, can be written as a product of $P_M(n)$ (see \eref{PMdef}) and an exponential function. The exponential will be a function of $n$, $M$, and a real parameter $\beta > 0$. For simplicity, we present the argument for the identity character with $M > 0$\footnote{The reasoning can be extended to the non-identity character and to cases with $M < 0$ as well.}.

Let us take $n\sim \cO(M)$ and $n\leq 2M$. We start by writing $a_n^{(2)}$ which is the part of $a_n$ (as in \eref{recurexpand}) with terms involving only $g(k,2)$ factors and its products,
\begin{align}
    a_n^{(2)} &\equiv P_M(n)\left[1 + \sum_{k_1=1}^{n-1}g(k_1,2) + \sum_{k_1=1}^{n-3}\sum_{k_2=k_1+2}^{n-1}g(k_1,2)g(k_2,2)\right. \nonumber\\
    &+ \sum_{k_1=1}^{n-5}\sum_{k_2=k_1+2}^{n-3}\sum_{k_3=k_2+2}^{n-1}g(k_1,2)g(k_2,2)g(k_3,2) + \ldots \nonumber\\ 
    &\left. + \sum_{k_1=1}^{n-(2i-1)}\sum_{k_2=k_1+2}^{n-(2i-3)}\ldots\sum_{k_i=k_{i-1}+2}^{n-1}g(k_1,2)g(k_2,2)\ldots g(k_i,2) + \ldots \right. \nonumber\\
    &\left.+ \sum_{k_1=1}^{n-(2i_{\text{max}}-1)}\sum_{k_2=k_1+2}^{n-(2i_{\text{max}}-3)}\ldots\sum_{k_{i_{\text{max}}}=k_{i_{\text{max}}-1}+2}^{n-1} g(k_1,2)g(k_2,2)\ldots g(k_{i_{\text{max}}},2) \right]
\end{align}
with $i_{\text{max}}=\left\lfloor\frac{n}{2}\right\rfloor$. Thus, the above expression has $i_{\text{max}}+1$ terms. 

Now note that $a_n^{(2)}$ above is the leading part of $a_n$ at every order since it only contains $g(k,p)$ with $p=2$. Here we are omitting all $p\geq 3$ factors. Now we can estimate $a_n^{(2)}$ with:
\begin{align}\label{approx_an2}
    a_n^{\text{exp}} = P_M(n)\exp\left(\sum_{k_1=1}^{n-1}g(k_1,2)\right) \equiv P_M(n)\exp\left(s^{(2)}\right).
\end{align}
With this now let us compute quadratic error. At quadratic order, we define the error as:
\begin{align}\label{quad_approx}
    e^{(2)} \equiv \frac{(s^{(2)})^2}{2} - \sum_{k_1=1}^{n-3}\sum_{k_2=k_1+2}^{n-1}g(k_1,2)g(k_2,2),
\end{align}
which after a bit of algebra, becomes:
\begin{align}\label{quad_err}
    e^{(2)} = \frac{1}{2}\sum_{k=1}^{n-1}g(k,2)^2 + \sum_{k=1}^{n-2}g(k,2)g(k+1,2),
\end{align}
which shows that $e^{(2)}>0$.

From the inductive proof, we know that $P_M(n)$ as defined in \eref{PMdef} and $a_n$ have the same sign. Now let us define below,
\begin{align}\label{r_beta}
    \tilde{r}(\beta,n,M) \equiv \frac{\left(\frac{a_n}{P_M(n)}\right)-
    \exp\left(s^{(2)}-\beta e^{(2)}\right)}{\left(\frac{a_n}{P_M(n)}\right)}.
\end{align}
where $\beta \geq 0$ is an arbitrary real parameter for now \footnote{One way to understand $\beta$ is the following. Note that we are only considering the quadratic error at this point that is $e^{(2)}$. However, there will also be cubic and higher-order errors coming from exponentiating $s^{(2)}$ alone and comparing that expression to $a_n^{(2)}$. The true expression will be a result of exponentiating terms involving $g(k,3)$, $g(k,4)$, $\ldots$, etc., -- that is all the single sum terms and not just $g(k,2)$. All of these higher-order error effects can be absorbed into a {\it renormalised parameter} and that is precisely the role of $\beta$ in \eref{r_beta}.}.

Now let us take $n=\alpha M$ with some fixed $\alpha$ in the range: $1<\alpha\leq 2$ in \eref{r_beta}. Let us argue that $\tilde{r}(\beta, M)<0$ for $\beta=0$. We begin by claiming: $|a_n^{\text{exp}}|>|a_n^{(2)}|$. One way to see this is that when we expand the exponential in $a_n^{\text{exp}}$ then at each order the nested sums in $|a_n^{\text{exp}}|$ go from $1$ to $n-1$ but the upper limit of the nested sums in $a_n^{(2)}$ keep on decreasing by $2$ (as can be seen in \eref{quad_approx}). This is exactly why $e^{(2)}>0$. 

Furthermore, $a_n^{(2)}$ has a finite set of terms, that is, $i_{\text{max}}+1$ terms, to be precise. However, $a_n^{\text{exp}}$ has an infinite set of terms upon expanding the exponential. Thus, $|a_n^{\text{exp}}|>|a_n^{(2)}|$.

One might think including $a_n^{(3)}$ -- which contains terms $g(k,3)$ and its products; $a_n^{(4)},\ldots$, etc., may lead to $|a_n|>|a_n^\text{exp}|$ but this won't happen since we already know that these are sub-leading w.r.t. $a_n^{(2)}$ (see \eref{subl}). Hence, we have argued that: $|a_n^{\text{exp}}|>|a_n|$.

From the above discussion, we see that $\tilde{r}(\beta,M)<0$ for $\beta=0$. Now, note that, if we take a large enough $\beta>0$ then we can make $\tilde{r}(\beta,M)>0$. Since $\tilde{r}(\beta,M)$ for a fixed $M$ and as a function of $\beta$ is continuous in $\beta$, by the Intermediate Value Theorem, there will exist a $\beta\in(0,\infty)$, say $\beta^*>0$, such that $\tilde{r}(\beta^*,M) = 0$. Then, we will have \footnote{Numerically, we find that for $j=1$ and $n=2M-1$ with large $2\leq M\leq 200$, $\beta^*\in(0,1)$. We conclude this by noting that for $\beta=0$, $\tilde{r}(0,M)<0$ and $|\tilde{r}(0,M)|$ increases with $M$. Additionally, for $\beta=1$, $\tilde{r}(1,M)>0$ and $|\tilde{r}(1,M)|$ again increases with $M$.}:
\begin{equation}\label{expr_mag_Mg0}
    a_n = P_M(n)\exp\left(s^{(2)}-\beta^* e^{(2)}\right)
\end{equation}
with $n=\alpha M$ and for some fixed $\alpha$ in the range: $1<\alpha\leq 2$. Note that, here $a_n$ is exact and not approximate. Thus, \eref{expr_mag_Mg0} is the exponential expression in terms of $M$ and $\beta^*$ we were after. This shows that $a_n$ as given in \eref{recurexpand} nicely exponentiates and further confirms the fact that the sign of $a_n$ is same as the sign of $P_M(n)$ for $1\leq n\leq 2M$.

\bibliography{l0_quasi}
\bibliographystyle{JHEP}

\end{document}